\documentclass[pre, superscriptaddress, showpacs, twocolumn]{revtex4-1}  
\usepackage{amssymb}
\usepackage{amsmath}
\usepackage{stix}
\usepackage[colorlinks,linkcolor=blue,citecolor=blue,urlcolor=blue]{hyperref}
\usepackage{graphicx}
\usepackage[utf8]{inputenc}
\usepackage[T1]{fontenc}

\usepackage[dvipsnames]{xcolor}

\renewcommand{\vec}[1]{\mathbf{#1}}

\newcommand{\dd}{\mathrm{d}}
\newcommand{\en}{\varepsilon}
\newcommand{\iu}{\mathrm{i}}

\newcommand{\hc}{\hat{c}}
\mathchardef\mhyphen="2D

\begin{document}

\author{Michael Sch\"uler}
\affiliation{Department of Physics, University of Fribourg, 1700 Fribourg, Switzerland}
\author{Martin Eckstein}
\affiliation{Department of Physics, University of Erlangen-N\"urnberg, 91058 Erlangen, Germany}
\author{Philipp Werner}
\affiliation{Department of Physics, University of Fribourg, 1700 Fribourg, Switzerland}

\title{Truncating the memory time in nonequilibrium DMFT calculations}

\date{\today}

\hyphenation{}

\begin{abstract} 
  The nonequilibrium Green's functions (NEGF) approach is a versatile 
  theoretical tool, which allows to describe the electronic structure,
  spectroscopy and dynamics of strongly correlated systems. The applicability 
  of this method is, however, limited by its considerable computational
  cost. Due to the treatment of the full two-time dependence
  of the NEGF the underlying equations of motion involve a 
  long-lasting non-Markovian memory kernel that results in at least a $N^3_t$
  scaling in the number of time points $N_t$. The system's memory time is,
  however, reduced in the presence of a thermalizing
  bath. In particular, dynamical mean-field theory (DMFT) -- one of the most
  successful approaches to strongly correlated lattice systems --
  maps extended systems to an effective impurity coupled to a bath. In this
  work, we systematically investigate how the memory time can
  be truncated in nonequilibrium DMFT simulations of the Hubbard and
  Hubbard-Holstein model. We show that suitable truncation schemes, which 
  substantially reduce the computational cost,  result in excellent approximations 
  to the full time evolution. This approach enables the
  propagation to longer times, making fundamental processes
  like prethermalization and the final stages of thermalization accessible
  to nonequilibrium DMFT.
\end{abstract}


\maketitle

\section{Introduction}

\begin{figure}[t] 
\centering
\includegraphics[width=\columnwidth]{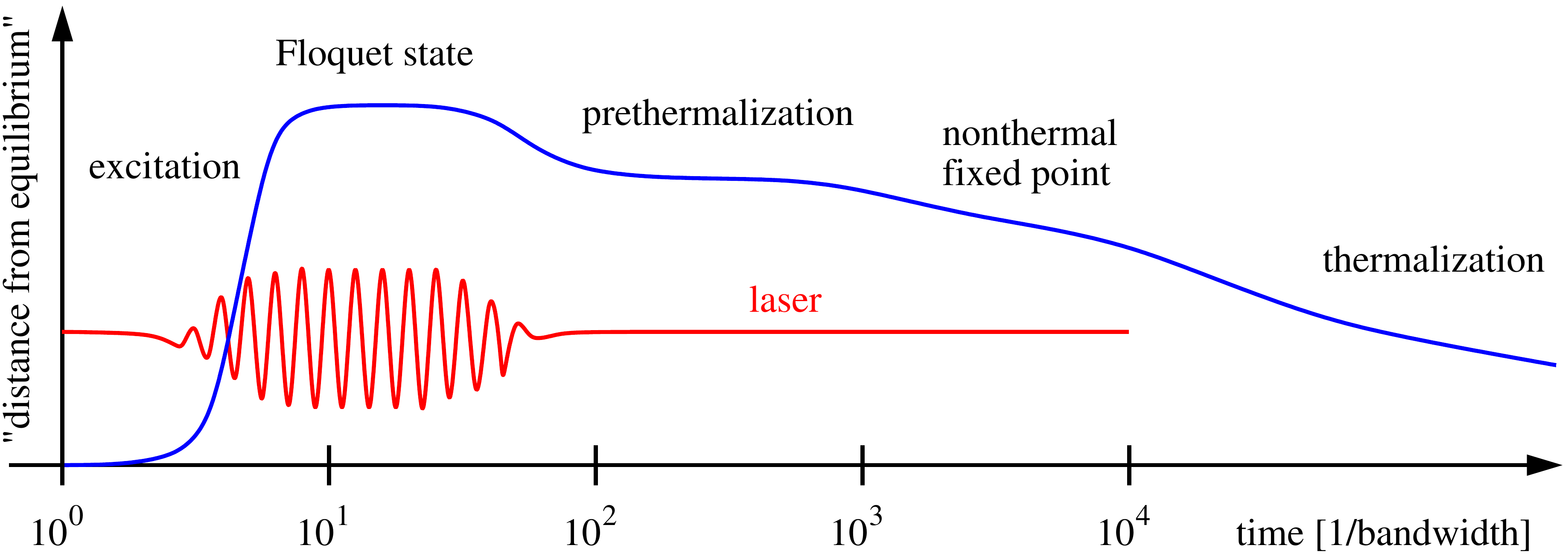}
\caption{Schematic
      illustration of the different timescales appearing in the
      evolution of a photo-excited correlated electron system. (Adapted
      from Ref.~\onlinecite{aoki_nonequilibrium_2014}.) }
\label{fig:timescales}
\end{figure}

The theoretical study of nonequilibrium phenomena in correlated
lattice systems is an active field of research, which is driven by the
rapid development of ultra-fast laser techniques and remarkable
experimental discoveries in light-driven
materials~\cite{fausti_light-induced_2011,kaiser_optically_2014,
  stojchevska_ultrafast_2014,mitrano_possible_2016,mor_ultrafast_2017}.
One of the challenges in the numerical simulation of the dynamics
after a photo-excitation or parameter quench is the emergence of
different relevant time scales (Fig.~\ref{fig:timescales}). In
lattice systems, 
the excitation process typically happens on the
timescale of the inverse electron hopping, corresponding to
femtoseconds in correlated electron
materials~\cite{cavalieri_attosecond_2007,krausz_attosecond_2009}. If
the driving laser field contains several cycles, the system may be
transiently described by a so-called Floquet state, which can exhibit
properties quite different from the equilibrium states of the initial
Hamiltonian. After the pulse, a relaxation process sets in, which eventually results in a new thermal
state. This relaxation may involve the transient trapping in
long-lived prethermalized
states~\cite{berges_prethermalization_2004,moeckel_interaction_2008,eckstein_thermalization_2009,marcuzzi_prethermalization_2013},
in which local observables look thermalized, while nonlocal ones are not, or
the passage near nonthermal critical
points~\cite{tsuji_nonthermal_2013,schuler_nonthermal_2018}. The
coupling to slow degrees of freedom such as phonons can introduce
additional timescales in the thermalization process 
\cite{golez_relaxation_2012,sentef_examining_2013,werner_field-induced_2015,murakami_interaction_2015}. %and ultimately lead to complete thermalization of the systems characterized by an effective temperature.

Developing analytical or numerical methods that can bridge these
different timescales and describe the evolution of the system from the
initial excitation process to the final, thermalized equilibrium state
is a major challenge. While numerical approaches such as time-dependent
exact diagonalization \cite{golez_relaxation_2012,lu_enhanced_2012},
density-matrix renormalization group methods~\cite{daley_time-dependent_2004} or the
nonequilibrium Green's functions (NEGF)
approach~\cite{balzer_nonequilibrium_2012,stefanucci_nonequilibrium_2013}
accurately capture the short-time evolution during the excitation and
initial relaxation process, a quantum kinetic description like the
Generalized Kadanoff-Baym Ansatz
(GKBA)~\cite{lipavsky_generalized_1986,
  pal_conserving_2009,latini_charge_2014,balzer_stopping_2016} or
related methods~\cite{galperin_linear_2008,
  ness_nonequilibrium_2011,bonitz_quantum_2015} allow to approximately describe the
thermalization dynamics. In
particular, the GKBA assumes a decoupling of the spectral properties
(encoded in the two-time spectral function) and the occupation
dynamics described by the single-particle reduced density matrix.  One
possible strategy is the development of a multi-scale approach, where
the photoexcited charge carrier distribution obtained by an accurate
method is used to initialize a kinetic equation. However, it is not
clear whether the ad-hoc approximations entailed by such an approach
compromise the long-time dynamics. This issue is particularly relevant
for systems with long-range order, where a fully self-consistent
treatment is crucial for the dynamics of the order parameter. For
these reasons, a formalism which allows to treat the entire evolution
and the relevant physical processes within a numerically efficient,
consistent and controllable scheme would be highly desired.

An approximate method which can, in principle, capture the different
stages in the time-evolution of a photo-excited correlated lattice
model is the dynamical mean field theory (DMFT)
\cite{georges_dynamical_1996,
  freericks_nonequilibrium_2006,aoki_nonequilibrium_2014}. It maps the
lattice system onto a self-consistently determined quantum impurity
model, and is formulated directly in the thermodynamic limit. While
the basic approximation of this scheme, the local nature of the
self-energy
\cite{metzner_correlated_1989,muller-hartmann_correlated_1989},
neglects processes that are related to nonlocal fluctuations and
correlations, it can describe the excitation by strong laser fields
\cite{freericks_nonequilibrium_2006,tsuji_nonequilibrium_2009,eckstein_thermalization_2011}, the
trapping in prethermalized states
\cite{eckstein_thermalization_2009,tsuji_nonthermal_2013}, and the relaxation
into a new thermal equilibrium state
\cite{eckstein_thermalization_2011,werner_role_2014}. Furthermore, the
method can be systematically extended to include nonlocal effects:
cluster versions of DMFT
\cite{maier_quantum_2005,tsuji_nonequilibrium_2014,eckstein_ultra-fast_2016}
allow to capture the effect of short-range correlations, while
extended DMFT
(EDMFT)~\cite{chitra_effective-action_2001,golez_dynamics_2015} and
combinations with the $GW$ approximation ($GW$+EDMFT) allow to
incorporate local and nonlocal polarization
effects~\cite{nilsson_multitier_2017,golez_nonequilibrium_2017-1}. This
versatility makes the nonequilibrium DMFT one of the currently most
powerful techniques for the study of photo-excited lattice models and,
in particular, strongly correlated systems.

\begin{figure}[b] 
\centering
\includegraphics[width=0.8\columnwidth]{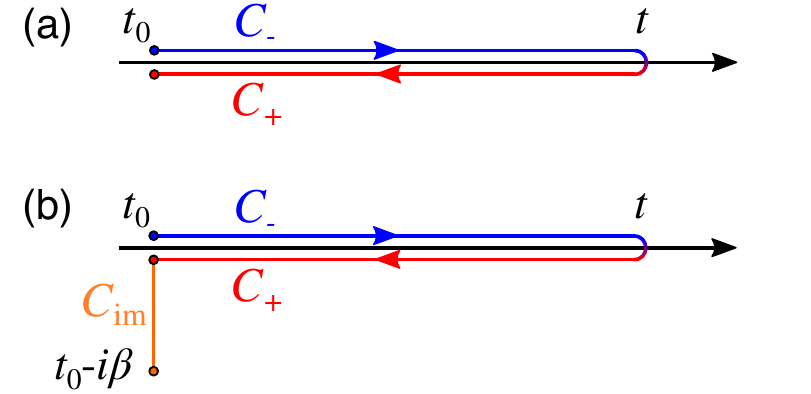}
\caption{The time arguments of the Green's functions lie on the
  contour $\mathcal{C}$ consisting of the forward ($\mathcal{C}_-$)
  and backward ($\mathcal{C}_+$) branches along the real axis, and the
  imaginary branch $\mathcal{C}_\mathrm{im}$. The arrows indicate the
  direction of the contour-ordering. $\beta$ denotes the inverse
  temperature. }
\label{fig:contours}
\end{figure}

A crucial step in this NEGF-based technique is the solution of the
two-time equation of motion for the single-particle Green's function (GF),
which defines an integro-differential equation on the Keldysh
(Fig.~\ref{fig:contours}(a)) or Kadanoff-Baym
(Fig.~\ref{fig:contours}(b)) contour. Many-body effects are captured
by the two-time self-energy, which plays the role of a memory kernel for the
time evolution. Even if a numerically cheap approximate impurity
solver is employed, nonequilibrium DMFT calculations implemented with
a fixed discretization of the time contour scale at least cubically
%(or even higher if higher-order impurity solvers are employed) 
in the number of time steps $N_t$.  (If a higher-order perturbative
impurity solver is employed, the numerical effort scales with a
correspondingly higher power of $N_t$
\cite{eckstein_nonequilibrium_2010,tsuji_nonequilibrium_2013}).  This
high computational cost originates from the fact that the entire
memory of the previous time evolution is kept in the calculation of
the interacting GF, which limits in practice the maximum
simulation times and thus the parameter regimes in which the full time
evolution from excitation to thermalization can be studied.

A very simple idea to overcome this limitation is to truncate the
memory time of the self-energy kernels. This approach roots in the
effective reduction of the system's memory when it is coupled to a
bath with broad spectrum, which is also the basis for Markovian
approximations in open quantum
systems~\cite{breuer_theory_2002,weiss_quantum_2012}. In the context
of DMFT, the mapping of the correlated electrons on a lattice to an
effective impurity model coupled to a (self-consistently determined)
bath should naturally result in a decay of the memory kernel. 
Hence, a properly implemented truncation scheme is expected to
have a small effect on the accuracy of the results. By introducing a
variable cutoff parameter, such an approach also allows to
systematically check and control the effect of the approximation. A
truncation of the memory time reduces the numerical cost of the Dyson
equation and related convolution integrals 
by at least one order, which should enable the study of
long-time dynamics without resorting to multi-scale approaches or
uncontrolled approximations. 

Motivated by this perspective, we investigate the effect of 
memory truncations in nonequilibrium DMFT. In particular, we
demonstrate the feasibility of the approach for paradigmatic examples:
the Hubbard model in the paramagnetic and antiferromagnetic (AFM) phase, as well
as for the Hubbard-Holstein model. The formalism is introduced and explained in
Sec.~\ref{sec:formalism}, while an analysis of the nonequilibrium
dynamics of these models for different memory times is presented and 
discussed in Sec.~\ref{sec:results}. Conclusions are presented in Sec.~\ref{sec:conclusions}.

\section{Formalism\label{sec:formalism}}

We focus our analysis on the Hubbard-Holstein model,
given by the Hamiltonian 
\begin{align}
  \label{eq:hubbhol}
  \hat{H} &= -v \sum_{\langle ij \rangle,\sigma} \hat{c}^\dagger_{i\sigma}
  \hat{c}_{j\sigma} 
  +\frac{U}{2}
  \sum_{i,\sigma}
  \hat{n}_{i\sigma}\hat{n}_{i\bar{\sigma}}  \nonumber \\ &\quad+ g\sum_{i,\sigma}
  \hat{n}_{i\sigma} (\hat{b}_i + \hat{b}^\dagger_i) 
          + \omega_0\sum_i \hat{b}^\dagger_i \hat{b}_i \ .
\end{align}
The first term describes the nearest-neighbor hopping of
electrons with spin $\sigma$ (fermionic creation and annihilation operators
$\hat{c}^\dagger_{i \sigma}$ and $\hat{c}_{i \sigma}$, respectively) with
amplitude $v$, the second term the on-site Coulomb repulsion of
electrons with different spin, parameterized by $U$, while the third term represents the
Holstein-type electron-phonon (e--ph) coupling (bosonic operators
$\hat{b}_i$). The last term is the phonon Hamiltonian. In what follows,
we measure energies in units of the hopping $v$, while the natural
time scale is given by $v^{-1}$. 

The central quantity in a NEGF-based treatment of the
Hamiltonian~\eqref{eq:hubbhol} is the single-particle lattice GF in real space
\begin{align}
  \label{eq:spgf}
  G_{ij,\sigma}(t,t^\prime) = - \iu \langle\mathcal{T}
  \hc_{i\sigma}(t) \hc^\dagger_{j\sigma}(t^\prime) \rangle \ , 
\end{align}
or in momentum space
\begin{align}
  \label{eq:spgfk}
  G_{\vec{k},\sigma}(t,t^\prime) = - \iu \langle\mathcal{T}
  \hc_{\vec{k}\sigma}(t) \hc^\dagger_{\vec{k}\sigma}(t^\prime) \rangle
  \ .
\end{align}
The time arguments of the GF are located on the Kadanoff-Baym contour
(Fig.~\ref{fig:contours}(b)), while $\mathcal{T}$ denotes the
corresponding contour-ordering operator. The operators and,
accordingly, the GF in momentum space are obtained by Fourier
transformation with respect to the underlying lattice. The latter also
defines the free electron dispersion $\en_{\vec{k}}$, which corresponds to the
eigenvalues of the hopping matrix in Eq.~\eqref{eq:hubbhol}. 

The lattice GF~\eqref{eq:spgfk} obeys the Kadanoff-Baym equation (KBE)
\begin{align}
  \label{eq:lattdyson}
  \left(\iu \partial_t + \mu -\en_{\vec{k}}\right)
  G_{\vec{k}\sigma}(t,t^\prime) - \int_\mathcal{C}\dd \bar{t} \, \Sigma_{\vec{k}\sigma}(t,\bar{t})
  G_{\vec{k}\sigma}(\bar{t},t^\prime) = \delta_C(t,t^\prime)
 \ . 
\end{align}
Here, $\mu$ stands for the chemical potential and the self-energy
$\Sigma_{\vec{k}\sigma}(t,t^\prime)$ (which is a functional of the GF)
captures all many-body effects. 

While the following analysis is carried out for the Hubbard and 
Hubbard-Holstein model, we stress that the statements on the effects of 
truncations of the memory time are generically valid. They also 
apply to DMFT simulations of multi-band models and extended DMFT formalisms.

\subsection{Dynamical mean-field theory in the strong-coupling limit}

DMFT maps a correlated lattice model (the
Hamiltonian~\eqref{eq:hubbhol} in our case) to a quantum impurity
model with a self-consistently determined bath. The main approximation
is the assumption of a spatially local self-energy, implying
$\Sigma_{\vec{k}\sigma}(t,t^\prime) \approx
\Sigma_{\sigma}(t,t^\prime)$. This approximation becomes
exact in infinite dimensional systems
\cite{metzner_correlated_1989,muller-hartmann_correlated_1989}.
 The local self-energy
$\Sigma_{\sigma}(t,t^\prime)$ can be computed from the solution of a
suitably defined auxiliary impurity system. Given a self-energy, the
lattice KBE~\eqref{eq:lattdyson} can be solved to obtain the approximate DMFT
lattice GF. The self-consistent solution is constructed
such that the local lattice GF,
\begin{align}
  \label{eq:gloc}
  G_\mathrm{loc,\sigma}(t,t^\prime) = \frac{1}{V_\mathrm{BZ}}
  \int_\mathrm{BZ} \! \dd \vec{k} \, G_{\vec{k}\sigma}(t,t^\prime)  \ ,
\end{align}
 is identical to the impurity Green's
function $G_\mathrm{imp,\sigma}(t,t^\prime)$.
Here BZ stands for the Brillouin zone and $V_\mathrm{BZ}$ the
corresponding volume.

Different methods can be employed to solve the impurity
problem~\cite{aoki_nonequilibrium_2014}. For the nonequilibrium scenarios
that we consider in this study, suitable methods for strong
electron-electron interaction are strong-coupling perturbative methods \cite{eckstein_nonequilibrium_2010}
such as the non-crossing approximation
(NCA)~\cite{keiter_diagrammatic_1971} or one-crossing approximation
(OCA)~\cite{pruschke_anderson_1989}. Here, the impurity problem
is treated by solving the local many-body problem exactly (energies
$E_\alpha$), which serves as a reference, while the hopping from and to
the surrounding bath is captured by the hybridization function
$\Delta_\sigma(t,t^\prime)$ (which is similar to an embedding
self-energy). Defining the so-called pseudo-particle GF
$\mathcal{G}_{\alpha}(t,t^\prime)$ ($\alpha$ labels the local
many-body states) as a correlator of the local many-body operators,
the impurity problem can be treated by diagrammatic methods. In
this case, the pseudo-particle self-energy $\Sigma_\alpha(t,t^\prime)$
becomes a functional of $\Delta_\sigma(t,t^\prime)$ and
$\mathcal{G}_{\alpha}(t,t^\prime)$. Expanding
$\Sigma_\alpha(t,t^\prime)$ in powers of the hybridization function,
the NCA corresponds to the first-order approximation beyond the atomic
limit, while OCA corresponds to the second-order scheme. 

The impurity KBE for the pseudo-particle GF $\mathcal{G}_\alpha$ is
similar to Eq.~\eqref{eq:lattdyson}. However, the convolution integral
has a slightly different form: the integrand is nonzero only if the
times $t'$, $\bar t$ and $t$ are in cyclic order on the contour (for
more details see Ref.~\onlinecite{eckstein_nonequilibrium_2010}),
\begin{align}
  \label{eq:impurity_dyson}
  \left(i\partial_t+\lambda-E_\alpha\right)\mathcal{G}_\alpha(t,t')-\int_{\mathcal{C},\text{cycl.}}
  \dd\bar{t}\, \Sigma_{\alpha}(t,\bar t)\mathcal{G}_\alpha(\bar t,
  t')=\delta_\mathcal{C}(t,t') \ .
\end{align}
Here, $\lambda$ denotes the pseudo-particle chemical potential.  Apart
from the difference in the convolution integral, the lattice and
impurity KBEs, Eqs.~\eqref{eq:lattdyson} and \eqref{eq:impurity_dyson},
respectively, have the same mathematical structure, and the same
scaling of the computational effort with the number of time steps. In the following, we will 
discuss a generic truncation scheme that  
applies to both cases.

\subsection{Solution scheme and memory cutoff\label{subsec:solcut}}

In this subsection, we briefly discuss the standard procedure for solving
a generic KBE. The goal is to introduce a consistent way of
truncating the memory.

\begin{figure}[t] 
\centering
\includegraphics[width=\columnwidth]{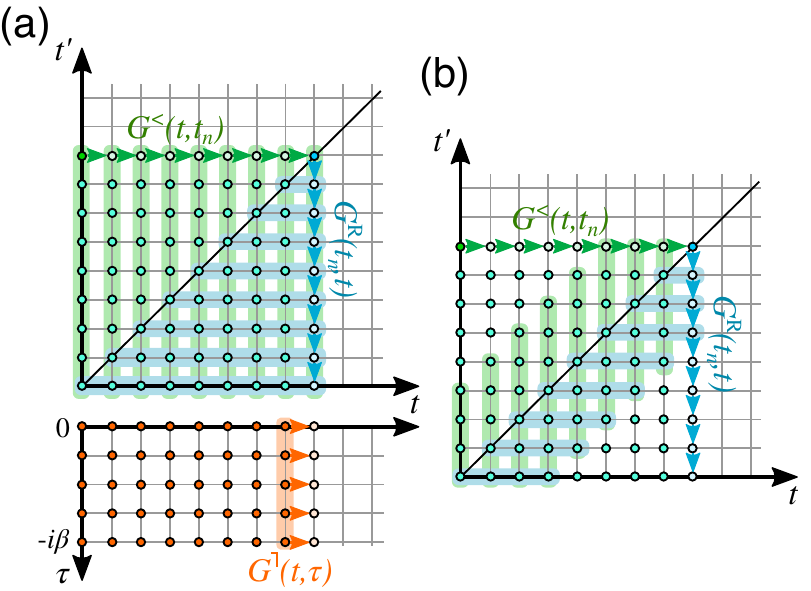}
\caption{(a) Sketch of the propagation scheme of the KBEs~\eqref{eq:genkbe} for the retarded and
  lesser GFs in the two-time plane (upper part) 
  and of the left-mixing
  GF (lower part). The method starts with the known value of
  $G^\mathrm{R}(t,t)$ at the diagonal, while successive time stepping
  (blue arrows) yields $G^\mathrm{R}(t,t^\prime)$ for all $t^\prime$
  up to $t$. Once $G^\mathrm{R}(t,0)$ is known, the KBE for $G^\rceil(t,\tau)$
  can be solved. Finally, using $G^<(0,t) =- [G^\rceil(t,0)]^\dagger$,
  the lesser GF can be propagated up to $G^<(t,t)$. 
  %The shaded slices in the background show the memory dependence of the corresponding self-energy for
  % carrying out one time step.
  The shaded lattice points indicate the values of the {\it self-energy} which are needed to calculate one time step.
  (b) Simplified propagation scheme (analogous to (a)) for times $t$ larger 
  than the cutoff $t_\mathrm{cut}$.}
\label{fig:propscheme}
\end{figure}

Suppose we are dealing with the generic KBE along with its adjoint version
\begin{subequations}
  \label{eq:genkbe}
  \begin{equation}
    \left(\iu \partial_t - \en \right)G(t,t^\prime) - I(t,t^\prime) =
    \delta_{\mathcal{C}}(t,t^\prime) \ ,
  \end{equation}
  \begin{equation}
    \label{eq:adjkbe}
    \left(-\iu \partial_t - \en \right)G(t^\prime,t) - \bar{I}(t^\prime,t) =
    \delta_{\mathcal{C}}(t,t^\prime) \ ,
  \end{equation}
\end{subequations}
where $I(t,t^\prime)$ [$\bar{I}(t,t^\prime)$] represents a convolution
integral of a generic self-energy $\Sigma(t,t^\prime)$ with
$G(t,t^\prime)$ [$G(t,t^\prime)$ with $\Sigma(t,t^\prime)$] as in
Eq.~\eqref{eq:lattdyson} or \eqref{eq:impurity_dyson}. The first step
in solving Eq.~\eqref{eq:genkbe} is to project the contour time
arguments onto real ($t,t^\prime \in \mathcal{C}_{\pm}$) and
imaginary ($t = -\iu \tau \in \mathcal{C}_\mathrm{im}$)
arguments. The different combinations give rise to
the different Keldysh components of the GF. Our
solution scheme is based on the retarded ($G^\mathrm{R}(t,t^\prime)$),
lesser ($G^<(t,t^\prime)$), left-mixing ($G^\rceil(t,\tau)$) and
Matsubara ($G^\mathrm{M}(\tau)$) components. All other Keldysh
components can be obtained as linear combinations of the above. Convolutions can be 
expressed in terms of the Keldysh components by using the standard Langreth
rules~\cite{stefanucci_nonequilibrium_2013} for lattice-type KBEs or
the modified Langreth rules for pseudo-particle
GFs~\cite{eckstein_nonequilibrium_2010}.

Assuming that the equilibrium problem has been solved and the Matsubara
component $G^\mathrm{M}(\tau)$ is known, the time propagation of the
real-time and left-mixing components can be performed according to the
sketch in Fig.~\ref{fig:propscheme}. We assume an equidistant
discretization $t_n = n \Delta t$ of the real-time axis into $N_t$
steps of size $\Delta t$. 

To compute the retarded component $G^\mathrm{R}(t_n,t_j)$ for all
$j=0,\dots,n$ one can start from the diagonal $G^\mathrm{R}(t_n,t_n)$
which is known from the commutation relations of the creation and
annihilation operators. Using the adjoint KBE~\eqref{eq:adjkbe} and
the Langreth rules, the
retarded GF is propagated by
\begin{align}
  -\iu \partial_t G^\mathrm{R}(t_n,t) =  G^\mathrm{R}(t_n,t) \en +
  \int^{t_n}_{t}\!\dd \bar{t} \, G^\mathrm{R}(t_n,\bar{t})
  \Sigma^\mathrm{R}(\bar t,t) \ .
\end{align}
Hence, to obtain $G^\mathrm{R}(t_n,t_j)$, the retarded self-energy
$\Sigma^\mathrm{R}(t_i,t_j)$ is needed for $i=j,\dots,n$, {\it i.\,e.}, the lattice points marked by the 
blue shaded background in Fig.~\ref{fig:propscheme}(a). The self-energy
$\Sigma^\mathrm{R}(t_n,t_i)$ for $i=0,\dots,n$ is further needed to 
obtain the left-mixing component $G^\rceil(t_n,\tau)$ and the
lesser component $G^<(t_j,t_n)$. The propagation scheme for the latter
(green arrows in Fig.~\ref{fig:propscheme}(a)), starting from
$G^<(0,t_n)$ and progressing towards the diagonal $G^<(t_n,t_n)$,
additionally requires the lesser self-energy $\Sigma^<(t_j,t_i)$ for 
$i=0,\dots,n$ for the calculation of $G^<(t_j,t_n)$ (lattice points marked by the green shaded
background in Fig.~\ref{fig:propscheme}(a)).  

We note that the KBEs can also be solved with different propagation
schemes, as explained, for instance, in
Refs.~\onlinecite{stan_time_2009,schuler_time-dependent_2016,schlunzen_nonequilibrium_2016}. The
dependence of the memory on the self-energy is, however, the same.

The computational effort for the solution of the
KBEs~\eqref{eq:genkbe} scales like $\mathcal{O}(N_t^3)$ assuming that
the self-energy is known. This is the bottleneck in calculations based on simple impurity solvers, such as the NCA. 
At the OCA level, the evaluation of the pseudo-particle self-energy
involves two internal 
integrals over the contour $\mathcal{C}$, so
that in this case the calculation of $\Sigma_\alpha$ at a cost
$\mathcal{O}(N_t^4)$ dominates simulation. In either case, the computational scaling can be
reduced by at least one order if the memory time of the self-energy is
truncated to $|t-t'| \le t_\mathrm{cut}$. Physically, a truncation of
the memory occurs, e.\,g., when
the
system is coupled to a bath without particle exchange. For instance,
a bath with broad spectrum would lead to a long-time decay of the form
 $G^\mathrm{R}(t,t^\prime) \sim e^{-\eta(t-t^\prime)}$,
$G^<(t,t^\prime) \sim e^{-\eta|t-t^\prime|}$ and
$G^\rceil(t,\tau)\sim e^{-\eta t}$, and analogously for the
self-energy. A consistent way to introduce a numerical memory cutoff
is to mimick this behavior 
by replacing $\Sigma^\mathrm{R}(t,t^\prime) \rightarrow f(t-t^\prime)
\Sigma^\mathrm{R}(t,t^\prime)$, $\Sigma^<(t,t^\prime) \rightarrow f(|t-t^\prime|)
\Sigma^<(t,t^\prime)$ and $\Sigma^\rceil(t,\tau) \rightarrow f(t)
\Sigma^\rceil(t,\tau)$, with a generic cutoff function $f(t)$. Here, we
choose a Fermi-function-like cutoff 
\begin{align}
  \label{eq:fermicut}
  f(t) = \frac{1}{1+\exp[(t-t_0)/T_c]} \ ,
\end{align}
which interpolates between a hard cutoff ($T_c \rightarrow 0$) and an 
exponential decay ($T_c \rightarrow \infty$). The cutoff time
$t_\mathrm{cut}$ after which memory effects can be neglected is thus defined 
by $f(t_\mathrm{cut})$ falling below a specified threshold.

For %sufficiently large time step 
$t_n > t_\mathrm{cut}$, the two-time
propagation scheme of the KBEs~\eqref{eq:genkbe} simplifies
significantly (Fig.~\ref{fig:propscheme}(b)). First, the left-mixing component
$\Sigma^\rceil(t_n,\tau)$ can be omitted. For this reason, solving
for $G^\rceil(t_n,\tau)$ is not required anymore. Furthermore, the calculation of 
the real-time Keldysh components requires only the information of the retarded and lesser self-energy (blue or green
shaded background in Fig.~\ref{fig:propscheme}(b), respectively) on a reduced 
time interval  of length $t_\text{cut}$.

\subsection{Memory cutoff via hybridization function}

We stress that the cutoff scheme introduced in
Sec.~\ref{subsec:solcut} is general and applies to different
types of NEGF setups. To illustrate the effect, we focus in the
following on DMFT calculations in the strong-coupling limit and regard
the GFs and the self-energy in Subsection~\ref{subsec:solcut} as
pseudo-particle quantities. For simplicity, we will consider an
infinite-dimensional Bethe lattice and employ an NCA impurity
solver. In this case the DMFT self-consistency condition for bandwidth
$4v$ simplifies to
$\Delta_\sigma(t,t^\prime)=v^2G_{\mathrm{loc},\sigma}(t,t^\prime)$
\cite{aoki_nonequilibrium_2014}, so that an explicit solution of the
lattice KBE~\eqref{eq:lattdyson} is not needed. In this case, the
cutoff is most conveniently introduced on the level of the
hybridization function:
\begin{subequations}
  \label{eq:hybcut}
  \begin{equation}
    \Delta^\mathrm{R}_\sigma(t,t^\prime) \rightarrow f(t-t^\prime)
    \Delta^\mathrm{R}_\sigma(t,t^\prime) \ ,
  \end{equation}
  \begin{equation}
    \Delta^<_\sigma(t,t^\prime) \rightarrow f(|t-t^\prime|)
    \Delta^<_\sigma(t,t^\prime) \ , 
  \end{equation}
  \begin{equation}
    \Delta^\rceil_\sigma(t,\tau) \rightarrow f(t)
    \Delta^\rceil_\sigma(t,\tau) \ .
  \end{equation}
\end{subequations}
Since each pseudo-particle self-energy $\Sigma_\alpha(t,t')$ contains
a factor $\Delta_\sigma(t,t')$ or $\Delta_\sigma(t',t)$, this implies
a corresponding truncation of the pseudo-particle self-energies.  We
remark that reducing the two-time dependence of the hybridization
function would also yield a significant reduction of the computational cost
of evaluating the internal integrals for the OCA or higher
approximations. The effort would then grow only like a power of the cutoff
time $t_\mathrm{cut}$, rather than a power of the maximum simulation
time.

\begin{figure}[t] 
\centering
\includegraphics[width=\columnwidth]{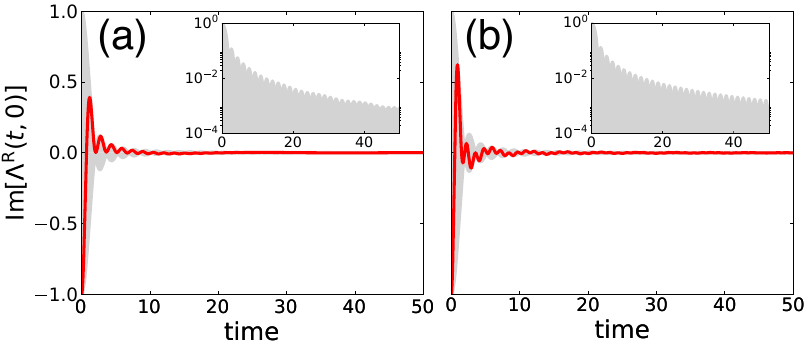}
\caption{Retarded component $\Lambda^\mathrm{R}_{\sigma}(t,0)$ of the
  hybridization function in equilibrium for (a) $U_0=4$, and (b)
  $U_0=6$. The gray shaded background represents the envelope function
  $2|\Lambda^<_{\sigma}(t,0)|$ (on a logarithmic scale in the inset).
  }
\label{fig:sbhb_gret_decay}
\end{figure}

\begin{figure*}[t] 
\centering
\includegraphics[width=0.9\textwidth]{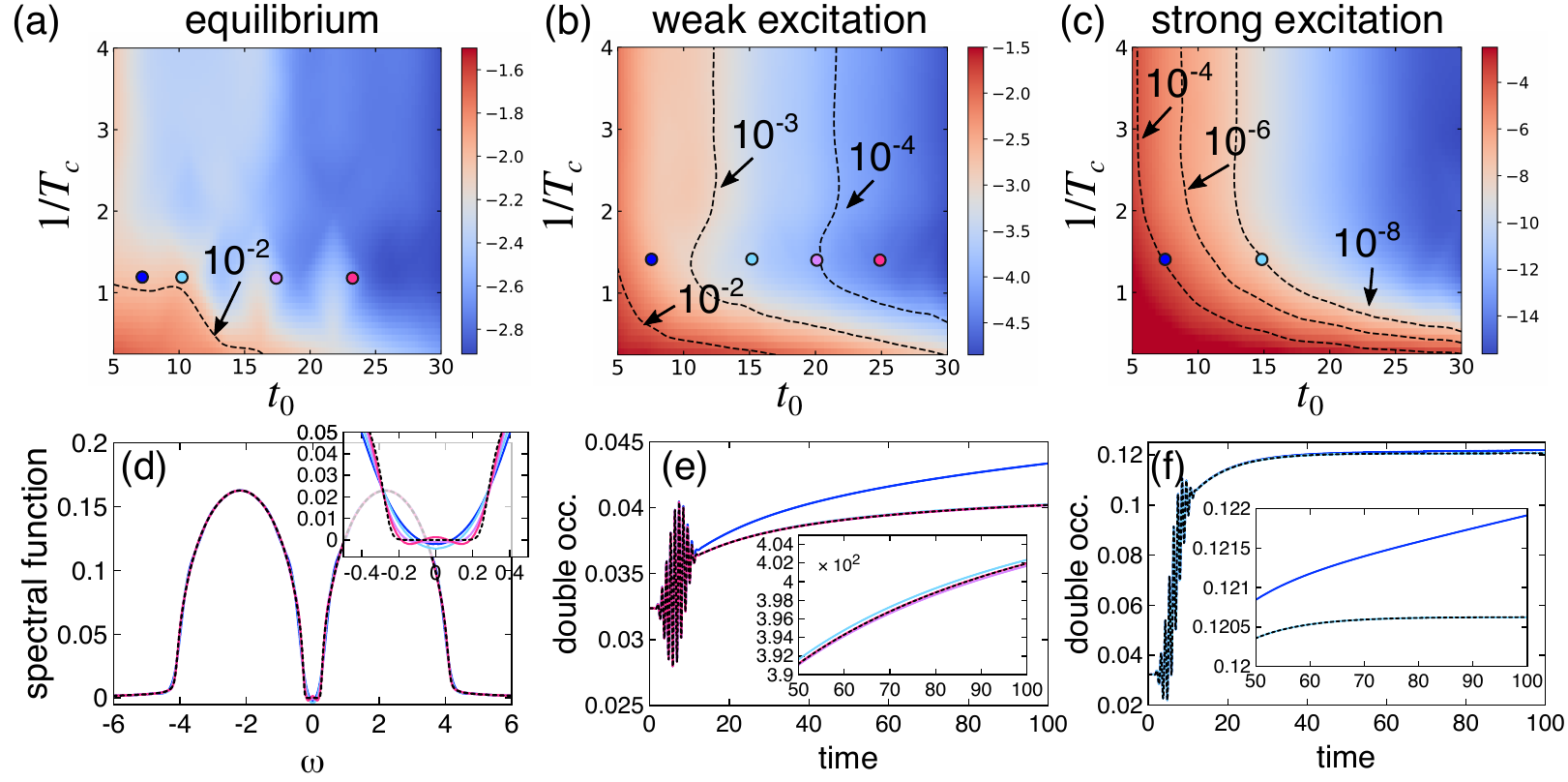}
\caption{(a)--(c): Norm distance (logarithm) between the two-time
  reference GF (no cutoff) and $G_\mathrm{cut}$ obtained by the cutoff
  scheme as a function of the memory time $t_0$ and sharpness of the
  cutoff $1/T_c$ (cf.~\eqref{eq:fermicut}) for $U_0=4$ and
  $\Delta U=0$ (a), $\Delta U =0.4$ (b) and $\Delta U=2.0$ (c). The
  contour lines delimit the regions where the error is smaller than
  the given values. The colored dots indicate representative values of
  $t_0$ and $T_c$, for which the equilibrium spectral function (d) and
  the double occupancy for weak (e) and strong excitation (d) are
  shown (consistent color coding). The black dashed lines represent
  the reference results.  
  }
\label{fig:sbhb_u4}
\end{figure*}

\section{Results\label{sec:results}}

\begin{figure}[t] 
\centering
\includegraphics[width=\columnwidth]{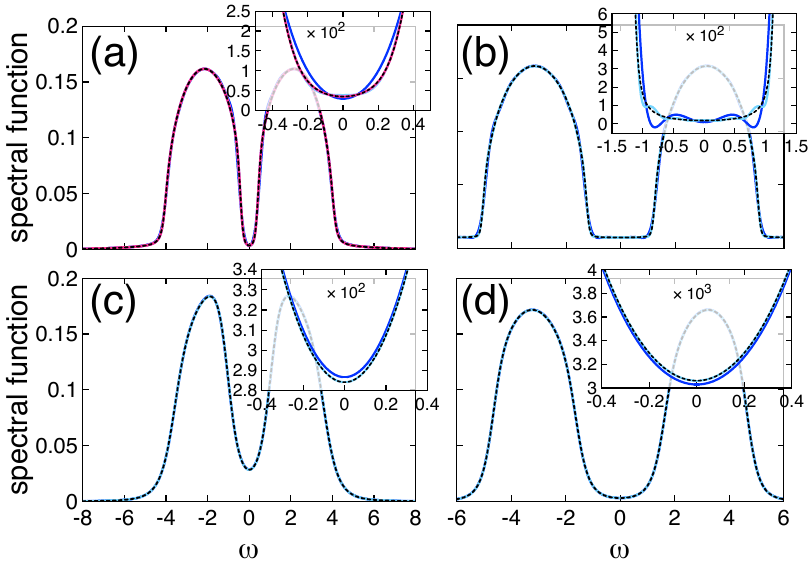}
\caption{Nonequilibrium spectral functions computed at
  $t_\mathrm{max}=100$ for (a) $U_0=4$, $\Delta U =0.4$, (b) $U_0=6$,
  $\Delta U = 0.6$, (c) $U=4$, $\Delta U = 2.0$, and (d) $U=6$,
  $\Delta U = 3$. Different lines correspond to different cutoff
  parameters $t_0$ with fixed $T_c$. The color coding corresponds to
  the circle symbols shown in Fig.~\ref{fig:sbhb_u4}(b) (for panel (a)
  and (b)) and Fig.~\ref{fig:sbhb_u4}(c) (for panel (c) and(d)). The
  insets show a zoom of the gap region.  For
  the case of strong excitation ((c) and (d), respectively), the
  spectral functions obtained by the cutoff scheme are identical to
  the reference one up to $10^{-4}$. }
\label{fig:sbhb_neq_spectral}
\end{figure}

\subsection{Hubbard model - paramagnetic phase\label{subsec:para}}

We first consider the simple case of a Hubbard model in the
paramagnetic phase. Nonequilibrium DMFT studies of this model have
provided fundamental insights into the nonequilibrium properties of
strongly correlated electron systems, including dynamical phase
transitions~\cite{eckstein_thermalization_2009}, dielectric breakdown
\cite{eckstein_dielectric_2010}, impact ionization \cite{werner_role_2014}
and thermalization \cite{eckstein_thermalization_2011}. Here we drive
the system out of equilibrium by an interaction modulation of the form
\begin{equation}
U(t)=U_0+\Delta U \sin(\omega_0 t) f_\mathrm{p}(t),
\end{equation}
with amplitude $\Delta U$ and pulse envelope
$f_\mathrm{p}(t) = \sin^2(\omega_0 t/2N_c)$ for
$0 < t < 2 \pi N_c / \omega_0$.  This form represents an $N_c$-cycle
pulse. Similar excitations by modulating the Hubbard repulsion have
been realized experimentally~\cite{jordens_mott_2008} and studied
theoretically~\cite{kollath_modulation_2006,eckstein_nonequilibrium_2010,peronaci_resonant_2017}. Here,
we focus on quasi-resonant excitations using a pulse with $N_c=10$
cycles and $\omega_0=U_0+2$, corresponding to transitions to the
higher energy part of the upper Hubbard band. In the small gap regime,
this can lead to the production of additional doublon-holon pairs by
impact ionization~\cite{werner_role_2014}.

To quantify the effect of the memory cutoff \eqref{eq:fermicut} we
calculate the time evolution (i) without imposing additional
approximations, which yields the reference DMFT GF $G(t,t^\prime)$,
and (ii), using the cutoff scheme \eqref{eq:hybcut} for the hybridization
function, which yields the approximate GF
$G_\mathrm{cut}(t,t^\prime)$. The inverse temperature is set to
$\beta=10$. In order to assess the effect of the truncation of the
hybridization function, we calculate the norm of the difference
$\left\lVert G-G_\mathrm{cut}\right\lVert$ for different cutoff
parameters $t_0$, $T_c$ and a maximum simulation time
$t_\text{max}=100$. One can define the norm of a two-time GF
in different ways. We consider the difference on the last time
slice according to the formula 
\begin{align}
  \label{eq:diff1}
  \left\lVert G\right\lVert &= \frac{1}{t_\mathrm{max}}
  \int^{t_\mathrm{max}}_0\!\dd t\,|G^<(t_\mathrm{max},t)| \nonumber \\
  &\quad+ \frac{1}{t_\mathrm{max}}
  \int^{t_\mathrm{max}}_0\!\dd t\, |G^\mathrm{R}(t_\mathrm{max},t)|
    \nonumber \\ &\quad + \frac{1}{\beta} \int^{\beta}_0\!\dd\tau\,
                   |G^\rceil(t_\mathrm{max},\tau)| \ .
\end{align}
Since the GF at the last time step depends on all the
previous steps, the definition~\eqref{eq:diff1} provides a convenient
way of comparing full two-time GFs. We have also tested the norm 
\begin{align}
  \label{eq:diff2}
  \left\lVert G\right\lVert &= \frac{1}{t^2_\mathrm{max}}
  \int^{t_\mathrm{max}}_0\!\dd t \int^{t_\mathrm{max}}_0\!\dd t^\prime\,|G^<(t,t^\prime)| \nonumber \\
  &\quad+ \frac{1}{t^2_\mathrm{max}}
  \int^{t_\mathrm{max}}_0\!\dd t\int^t_0\!\dd t^\prime\, |G^\mathrm{R}(t,t^\prime)|
    \nonumber \\ &\quad + \frac{1}{t_\mathrm{max}\beta} \int^{t_\mathrm{max}}_0\!\dd t \int^{\beta}_0\!\dd\tau\,
                   |G^\rceil(t,\tau)| \ ,
\end{align}
and found the results to be qualitatively similar to those obtained by 
Eq.~\eqref{eq:diff1}. Since Eq.~\eqref{eq:diff1} is faster to
evaluate, we employ it in the following analysis. 

Before we study the dynamics in detail, let us consider the time
dependence of the equilibrium hybridization function
$\Lambda^\mathrm{R}_{\sigma}(t,0)\equiv \Lambda^\mathrm{R}(t,0)$,
shown in Fig.~\ref{fig:sbhb_gret_decay} along with a bounding envelope
function defined by $|\Lambda^\mathrm{R}_{\sigma}(t,0)| \leq
|\Lambda^>_{\sigma}(t,0)| + |\Lambda^<_{\sigma}(t,0)|$
($=2|\Lambda^<_{\sigma}(t,0)|$ in the particle-hole symmetric case).
As one infers from the figure, the retarded hybridization function
(and thus all other real-time components) decays rapidly. In the
presence of a Mott gap we can decompose
$|\Lambda^\mathrm{R}_{\sigma}(t,0)|$ into a lower (corresponding to
$\Lambda^<_{\sigma}(t,0)$) and upper Hubbard band (corresponding to
$\Lambda^>_{\sigma}(t,0)$). Both the lesser and greater component are defined by a Fourier
transform over a frequency domain with semi-infinite support (in
the limit of $\beta\rightarrow \infty$). The Paley-Wiener
theorem~\cite{pavlyukh_time_2013} then implies that the decay of the
hybridization function in real time is of the form
$\Lambda^\gtrless(t,0)_\sigma\sim \exp(-B t^\alpha)$ with $0 < \alpha <
1$. Although the temperature is finite in our study, a sub-exponential
decay of the envelope function is observed in
Fig.~\ref{fig:sbhb_gret_decay} due to the finite gap size and the
exponentially vanishing spectral density within the gap.
Nevertheless, since the pseudoparticle self-energy involves a
multiplication with the (typically algebraically) decaying pseudoparticle GF, truncating
the hybridization function at $t_0 \sim 30$ is expected to yield a
good approximation. In nonequilibrium calculations, the partial
filling of the gap will result in a usual exponential decay, and the
effective memory time will be further reduced.

As key quantities to assess the quality of the truncation
approximation we consider, besides the GF norm difference, also the  
(equilibrium or nonequilibrium) spectral functions obtained by the ``backward" Fourier
integration
%\textcolor{red}{[we usually integrate from $t$ to $t_{max}$, that is why I worte "forward". In your case it seems to be backward in time ...]}
\begin{align}
  A(\omega,t)=-\frac{1}{\pi}\mathrm{Im}\int^{t}_0\!\dd t^\prime
e^{i\omega(t-t^\prime)}G^\mathrm{R}(t,t^\prime) \ ,
\end{align}
and, as an example of a local observable, the time-dependent double
occupation $d(t)=\langle n_\uparrow(t)n_\downarrow(t)\rangle$. The
results for $U_0=4$ (small gap Mott insulator case) are shown in
Fig.~\ref{fig:sbhb_u4}, while those for $U_0=6$ (larger gap case) can
be found in the Appendix (Fig.~\ref{fig:sbhb_u6}). Panels (a), (b) and
(c) show the norm error in the Green's function for the equilibrium
system ($\Delta U=0$), a weak excitation pulse ($\Delta U=0.1 U_0$)
and a strong excitation pulse ($\Delta U=0.5 U_0$), respectively. The
color scale indicates the logarithm of the error, and the dashed
contour lines correspond to fixed values of the error in the plane of
$t_0$ and $1/T_c$.

First of all, we notice that in this paramagnetic Hubbard model simulation, 
a sharp memory cutoff (large value of
$1/T_c$) has no particularly detrimental effect on the accuracy of the
GF. Interestingly, however, the best approximations for
the equilibrium system and the weakly excited system are obtained for
a relatively well defined cutoff temperature $T_c$ in the range
$0.75-0.9$. As the excitation strength is increased, the optimal
cutoff temperature becomes lower and less well defined.

The equilibrium system exhibits the largest cutoff effects. To reduce
the norm error down to $10^{-3}$ one has to choose memory times
$t_0\approx 30$, consistent with the rough estimate 
based on Fig.~\ref{fig:sbhb_gret_decay}, 
and even this cutoff still produces small artifacts
in the spectral function. This is illustrated in panel (d) of
Figs.~\ref{fig:sbhb_u4} and \ref{fig:sbhb_u6}, which compares the
exact spectral function (dashed) to the approximate spectral functions
for the cutoff parameters indicated by the colored dots in panel
(a). In particular, the sharp band edges are not well reproduced (see
inset of panel (d)) and there appears some negative spectral weight in
the gap region, which is unphysical. The approximate spectral function
however approaches the exact result with increasing
$t_0$, which shows that the truncation errors can be systematically
controlled. Sharp spectral features are associated with slowly
decaying Green's functions, and this in turn implies slowly decaying
pseudo-particle Green's functions and self-energies. It is thus not
surprising that the equilibrium system at low temperature represents a
challenging test case for our cutoff scheme.

Simulations with truncated memory time are much more accurate in the
nonequilibrium case. Even after a weak excitation (panel (b) in
Fig.~\ref{fig:sbhb_u4} and \ref{fig:sbhb_u6}), errors in the GF of the
order $10^{-4}$ can be reached with $t_0$ in the range 15-30
(depending on $U$) and this level of accuracy is sufficient to produce
spectral functions and local observables that are, within the
numerical accuracy of the simulation, hardly distinguishable from the
results of the full calculations.  In panel (e) we plot the
time-evolution of the double occupation, for the cutoff parameters
indicated by the colored dots in panel (b). Already the light-blue
curve ($t_0=15$) is accurate to more than four digits (see inset), and
this error remains constant up to the longest simulation
times. Comparing the results of Figs.~\ref{fig:sbhb_u4}(e) and
\ref{fig:sbhb_u6}(e), we furthermore note that in the case of a small
gap insulator (Fig.~\ref{fig:sbhb_u4}) the thermalization dynamics
involves two timescales. The fast timescale (i.e. the dynamics up to
time $t\approx 40$) can be associated with doublon-holon creation by
impact ionization \cite{werner_role_2014}, while the longer
(thermalization) timescale is associated with doublon-holon creation
by multi-particle scattering. In the large-gap insulator
(Fig.~\ref{fig:sbhb_u6}), the impact ionization is suppressed, because the
kinetic energy of the pulse-induced doublons and holons is not
sufficient to produce additional doublon-holon pairs. The simulation
with memory cutoff correctly reproduces this physics.

Finally, in the strong pulse excitation case, where a large amount of
energy is injected into the system and the ``photo-doped" doublon
density reaches several percent, the memory time becomes very
short. In these simulations, $t_0\approx 10$ is sufficient to
reproduce the exact results with 5 digits accuracy, see panels (c) and
(f).

The results for the nonequilibrium spectral function, plotted in
Fig.~\ref{fig:sbhb_neq_spectral}, confirm the conclusions drawn from
the previous analysis. In the weak excitation case (panels (a) and
(c)) small artifacts appear in the gap region if the cutoff time is
chosen too small ($t_0=7.5$), while $t_0=15$ is sufficient to
reproduce the exact time-dependent spectra. In the case of a
strong excitation (panels (c) and (d)), even $t_0=7.5$ is adequate.

In general, stronger excitations result in a partial filling the Mott
gap as a result of heating and photo-doping. For this reason, the the
Paley-Wiener theorem does not apply any more, and the (nonequilibrium)
hybridization functions $\Lambda^\gtrless_\sigma(t,t^\prime)$ decay
exponentially in $|t-t^\prime|$. Hence, truncating the memory works
better for stronger excitations.

\begin{figure}[t] 
\centering
\includegraphics[width=\columnwidth]{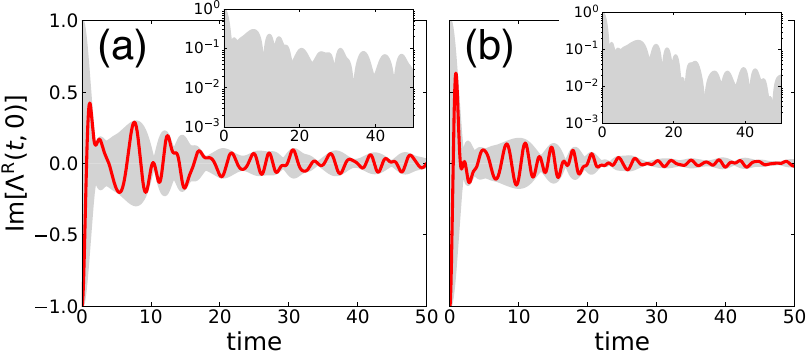}
\caption{Retarded component $\Lambda^\mathrm{R}_{\uparrow}(t,0)$ of the
  hybridization function in equilibrium (AFM case) for (a) $U_0=4$, and (b)
  $U_0=6$. The gray shaded background represents the envelope function
  $|\Lambda^>_{\uparrow}(t,0)| + |\Lambda^<_{\uparrow}(t,0)|$ (on a logarithmic scale in the inset).}
\label{fig:afm_gret_decay}
\end{figure}

\begin{figure*}[t] 
\centering
\includegraphics[width=0.9\textwidth]{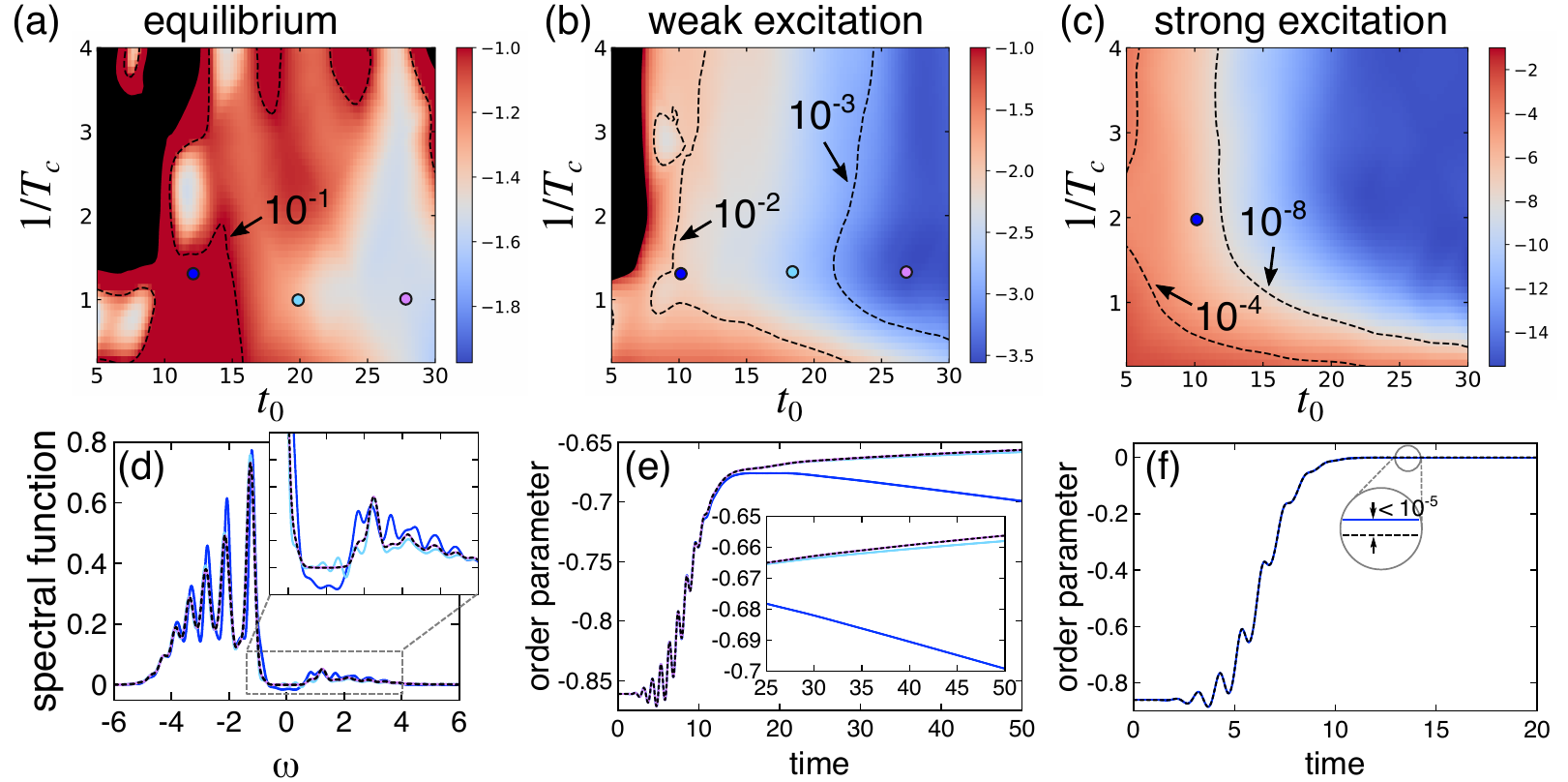}
\caption{(a)--(c): Norm distance (logarithm) between the two-time
  reference GF $G_{\mathrm{loc},\uparrow}(t,t^\prime)$ (no cutoff) and
  the GF obtained by the cutoff scheme, analogous to
  Fig.~\ref{fig:sbhb_u4}, for $U_0=4$ and $\Delta U=0$ (a),
  $\Delta U =0.4$ (b) and $\Delta U=2.0$ (c) in the AFM
  phase. The black region indicates cutoff parameters for which the 
  propagation with time step $\Delta t=0.01$ is unstable. (d) Equilibrium majority spin spectral function
  obtained from the reference GF (black dashed) and the cutoff scheme (colors as in
  (a)). The dynamics of the AFM order parameter
  $\langle \hat{n}_\uparrow \rangle - \langle \hat{n}_\downarrow
  \rangle$ is shown for $\Delta U = 0.4$ in panel (e), and for 
  $\Delta U=2.0$ in panel (f).  }
\label{fig:afm_u4}
\end{figure*}

\subsection{Hubbard model - antiferromagnetic phase}

As the next paradigmatic example we study the Hubbard model in the
AFM phase, which can be stabilized by the self-consistency condition 
$ \Delta_\sigma(t,t^\prime) = v^2
G_{\mathrm{loc},\bar\sigma}(t,t^\prime) $. We choose the same
excitations as in Sec.~\ref{subsec:para} and consider the 
equilibrium system, and the case of weak and strong perturbation. Here, a smaller time step of
$\Delta t =0.01$ is required for a stable propagation, hence we compute the DMFT solution 
up to $t_\mathrm{max} = 50$. 

We start again by analyzing the retarded component of the equilibrium hybridization function,
which yields the spectral function by Fourier transformation. The result, shown in Fig.~\ref{fig:afm_gret_decay}, 
is quite different from the paramagnetic case. Long-lived oscillations are
present, which give rise to characteristic sharp spectral features
associated with antiferromagnetic excitations. Inspecting
the decay of the envelope function for $U_0=4$
(Fig.~\ref{fig:afm_gret_decay}(a)), which stays above $10^{-2}$ up to $t_\mathrm{max}$, it appears that any
truncation of the memory should result in a quite poor 
approximation. This ``worst case" scenario
allows us to investigate the artifacts introduced by truncating the
memory at too early times. It turns out that a sufficiently large
cutoff time $< t_\text{max}$ nevertheless captures the main features and yields a good
approximation of the equilibrium properties and, in particular, the
nonequilibrium dynamics. 
The quality of the cutoff approximation can be 
expected to substantially improve for $U_0=6$
(Fig.~\ref{fig:afm_gret_decay}(b)), since in this case the hybridization function
approaches zero faster.

\begin{figure}[t] 
\centering
\includegraphics[width=0.8\columnwidth]{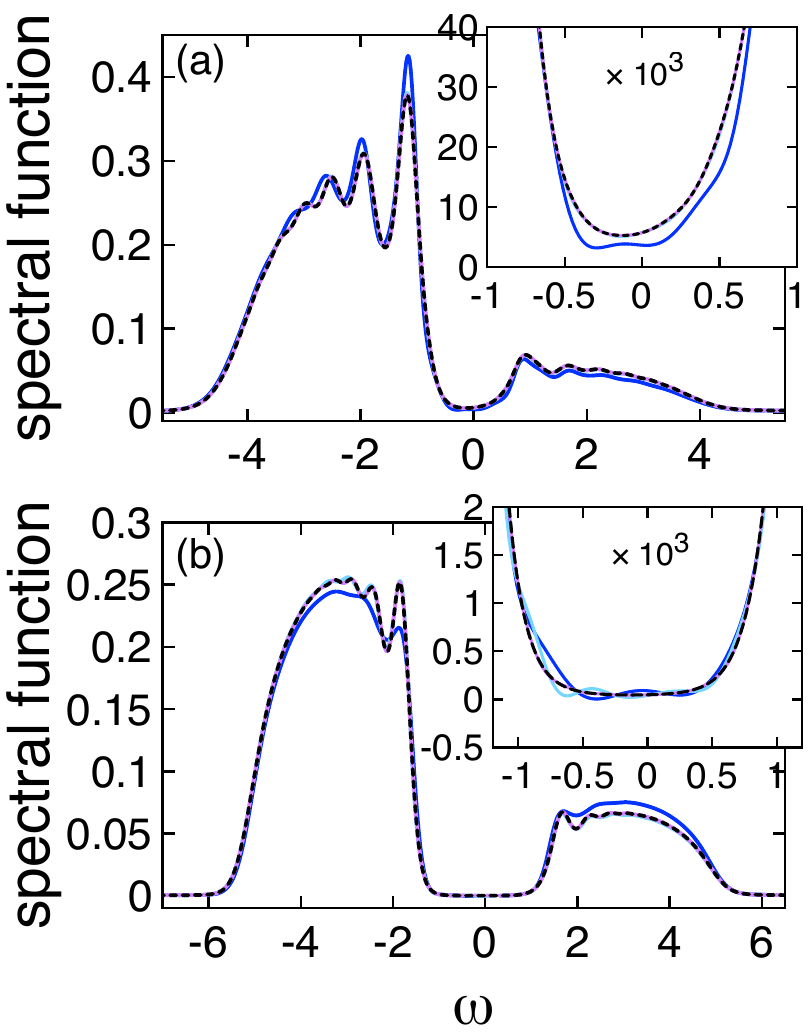}
\caption{Nonequilibrium (spin-up) spectral functions computed at
  $t_\mathrm{max}=50$ for (a) $U_0=4$, $\Delta U =0.4$, and (b)
  $U_0=6$, $\Delta U = 0.6$.  Different lines correspond to different
  cutoff parameters $t_0$ and $T_c$. The color coding corresponds to
  the circle symbols shown in Fig.~\ref{fig:afm_u4}(b) (for panel (a))
  and Fig.~\ref{fig:afm_u6}(b) (for panel (c).  }
\label{fig:afm_neq_spectral}
\end{figure}

Fig.~\ref{fig:afm_u4} presents an analysis of the small-gap system
analogous to Sec.~\ref{subsec:para}, while the results for
larger gap can be found in the Appendix (Fig.~\ref{fig:afm_u6}). In
the equilibrium case, as expected from the slow decay of the
hybridization function, the norm error of the GF produced by the
cutoff scheme is generally larger than for the paramagnetic case
(Fig.~\ref{fig:afm_u4}(a)-(c)). In particular, the error does not fall
below $10^{-2}$. 
Furthermore, the time propagation for
$\Delta t=0.01$ can become unstable for certain cutoff parameters 
($t_0$ and $T_c$ chosen in the black regions of the figure).
Inspecting the time-dependent observables, which
should be constant in equilibrium, we found that the total density
$\langle \hat{n}_\uparrow \rangle + \langle \hat{n}_\downarrow
\rangle$ is conserved up to an accuracy of $\sim 10^{-5}$ or better for
any values of the cutoff parameters. 
In contrast, the AFM order parameter
$\langle \hat{n}_\uparrow \rangle - \langle \hat{n}_\downarrow
\rangle$ violates the corresponding conservation law (depending on the cutoff
parameters). In particular, for too small values of $t_0$, the
magnitude of the AFM order decreases. This effect is most
pronounced in the unstable region. The instability is related to
negative spectral weight originating from a sharp cutoff ($T_c > 1.5$). In
fact, multiplying the exact reference hybridization function with the
cutoff according to Eq.~\eqref{eq:hybcut} and performing the Fourier
transformation, one obtains a spectral function with 
negative weight in the gap region. This breaks the conservation of
$\langle \hat{n}_\sigma \rangle$ and therefore of the order parameter. 

Inspecting the equilibrium spectral functions
(Fig.~\ref{fig:afm_u4}(d)) obtained by the time evolution with memory
cutoff, we find sharp spectral features originating from
antiferromagnetic excitations, which are the reason for the slow
decay of $\Delta_\sigma(t,t^\prime)$ and the difficulties of the cutoff
procedure in the equilibrium case. In particular, negative
spectral weight appears in the gap for $T_c = 0.83$ and $t_0=12$
(blue line). Increasing $t_0$, the cutoff results still deviate
substantially from the exact spectrum, unless the cutoff time $t_0$ is
increased up to $t_0\approx 27$.

The situation improves after a weak excitation
(Fig.~\ref{fig:afm_u4}(b)). Apart from an unstable region at small
$t_0$, the error can be reduced to less than $10^{-3}$ for cutoff
times $t_0 > 20$ for $1/T_c \approx 0.82$, while further increasing $t_0$
yields even smaller errors. As a relevant local observable, we compare
in Fig.~\ref{fig:afm_u4}(e) the AFM order parameter
obtained for different cutoffs. We
find that $t_0 \approx 20$ is sufficient to converge the dynamics of
the order parameter to an absolute deviation of less than $10^{-3}$,
even though in this excitation regime the order parameter is reduced by only about 20\%.
After a strong excitation (Fig.~\ref{fig:afm_u4}(c)), on the other
hand, the magnetic order melts rapidly, such that the behavior
discussed for the paramagnetic system in Sec.~\ref{subsec:para} is
recovered to a large extended. The evolution of the order parameter is
shown in Fig.~\ref{fig:afm_u4}(f) and deviates by less than $10^{-5}$
from the exact solution.

\begin{figure}[t] 
\centering
\includegraphics[width=\columnwidth]{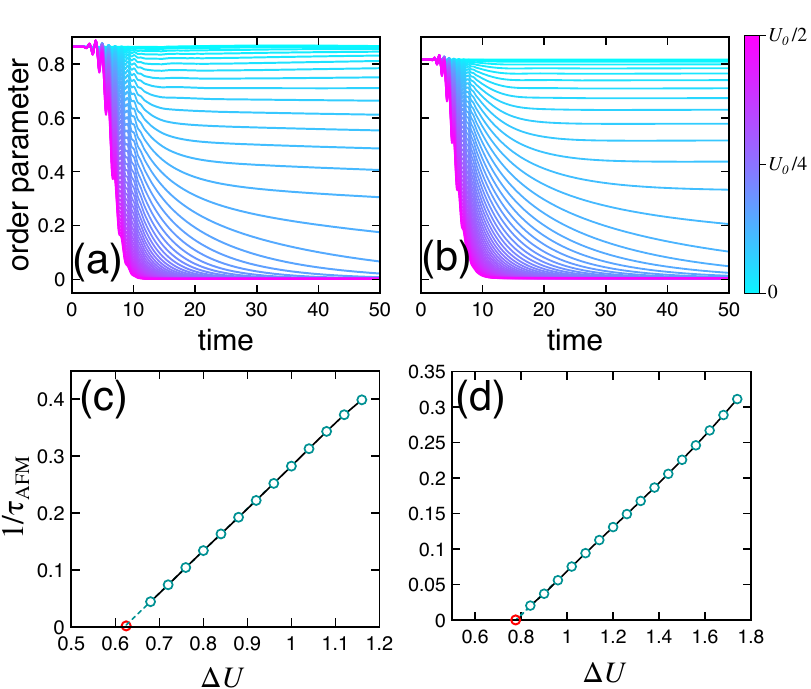}
\caption{Time-dependent AFM order parameter as a function of the
  excitation strength $\Delta U$ (color coding) for (a) $U_0=4$, and
  (b) $U_0=6$.  The inverse decay rate $\tau$ of the AFM order
  parameter obtained with $t_0=15$ and $T_c=0.8$ (circles) is shown in
  (c) for $U_0=4$ and in (d) for $U_0=6$. The black solid line
  shows the results obtained without any cutoff, while the
  dashed line indicates the linear extrapolation to the critical point
  where $1/\tau_\mathrm{AFM} = 0$ (marked by the red circle). }
\label{fig:afm_nonthermal}
\end{figure}

Increasing the Hubbard repulsion to $U_0=6$ (see Appendix, Fig.~\ref{fig:afm_u6}),
one finds a similar behavior as for $U_0=4$ except that the artifacts in the equilibrium
case (Fig.~\ref{fig:afm_u6}(a) and (d)) are less pronounced.  
The equilibrium time propagation with time step $\Delta t=0.01$ is
stable for $t_0>12$ for all $T_c$ and the error is reduced 
below $10^{-2}$ for $t_0>25$ and small enough $T_c$. Inspecting the
equilibrium spectral functions, good agreement is found for
$t_0 > 20$. This can be understood from the faster oscillations of the hybridization function
(corresponding to spectral features at larger $|\omega|$). If there are more oscillations in a given time
interval, the negative spectral weight arising from a
cutoff is reduced. Furthermore, the peaks of the spectral function are
less pronounced than for $U_0=4$. The error of the GF for weak
excitations (Fig.~\ref{fig:afm_u6}(b)) can be suppressed to less than
$10^{-3}$ by choosing $t_0> 15$. For $t_0>25$, the AFM order parameter
deviates by less $\sim 10^{-4}$ from the exact result. Again, the behavior after a strong excitation 
(Fig.~\ref{fig:afm_u6}(c) and (f)) resembles the strongly excited
paramagnetic case, where the memory of the hybridization functions
drops rapidly and leads to excellent approximations even for short cutoff times.

We have also computed the nonequilibrium spectral function at
$t=t_\mathrm{max}$ for the case of weak excitations
(Fig.~\ref{fig:afm_neq_spectral}). Due to the reduction of the AFM
order parameter, the sharp spectral features are washed out, and we find 
a good agreement with the reference spectral function for 
moderate values of $t_0$.  Nevertheless, for an accurate estimate of the 
the electronic gap, $t_0\approx 18$ is required.

In order to study nonequilibrium dynamics specifically associated with
symmetry-broken phases we perform a scan over $\Delta U$. Depending on
the excitation strength one observes two qualitatively different
behaviors of the AFM order parameter, and diverging timescales near
the critical excitation amplitude.  This is a manifestation of a {\it
  dynamical phase transition}. We note that the latter term has been
used in various
contexts~\cite{eckstein_thermalization_2009,schiro_time-dependent_2010,
  heyl_dynamical_2014,zunkovic_dynamical_2018} and that numerical
simulations have played an important role in revealing these
phenomena.  In the present study, the nonequilibrium transition is
between two long-lived nonthermal states with and without AFM order,
and the (almost) diverging timescale is associated with the melting of
the order, or the relaxation into the trapped AFM state
\cite{werner_nonthermal_2012}. The universal character of this type of
dynamical phase transition is exemplified by its appearance in a broad
range of systems, including
superconductors~\cite{yuzbashyan_dynamical_2006,barankov_synchronization_2006,peronaci_transient_2015,mazza_sudden_2017},
excitonic insulators~\cite{murakami_photoinduced_2017},
antiferromagnetic~\cite{werner_nonthermal_2012,tsuji_nonthermal_2013},
ferromagnetic~\cite{marcuzzi_prethermalization_2013,zunkovic_dynamical_2018,lerose_chaotic_2018}
and charge-ordered~\cite{schuler_nonthermal_2018} systems.
Fig.~\ref{fig:afm_nonthermal} shows the dynamics of the AFM order
parameter as a function of $\Delta U$ for both the small-gap
(Fig.~\ref{fig:afm_nonthermal}(a)) and the wide-gap case
(Fig.~\ref{fig:afm_nonthermal}(b)) with cutoff parameters $t_0=15$ and
$T_c=0.8$. The behavior is qualitatively similar for both values of
$U_0$: For $\Delta U < 0.1 U_0$, the order parameter exhibits
amplitude oscillations 
after the pulse ($t\gtrsim 10$)
%\textcolor{red}{[ARE THE OSCILLATIONS SEEN up to t=10 due to the pulse?]}
with a frequency related to the gap
size. With increasing excitation strength, the oscillation frequency
decreases and becomes difficult to measure near the 
nonthermal critical point. Testing different values
of the cutoff parameters we found that an accurate description of the regime of amplitude mode
oscillations requires $t_0 \approx 25$, especially
for $U_0=4$. This is consistent with the error analysis in
Fig.~\ref{fig:afm_u4}. In contrast, the regime of stronger
excitations -- above the excitation threshold for nonthermal melting of the AFM -- is well captured by a
memory time as short as $t_0=15$. As can be seen in Fig.~\ref{fig:afm_nonthermal}, a strong
excitation $\Delta U \approx 0.5 U_0$ results in a rapid melting of
the AFM order. Decreasing $\Delta U$ one approaches the nonthermal
critical point and the melting time increases. An 
exponential fit allows us to extract the characteristic timescale 
$\tau_\mathrm{AFM}$, whose inverse $\tau^{-1}_\mathrm{AFM}$ is plotted in
Fig.~\ref{fig:afm_nonthermal}(c)--(d). One finds a linear behavior of
$\tau^{-1}_\mathrm{AFM}$ as a function of $\Delta U$, in agreement with Ref.~\cite{werner_nonthermal_2012}. 
The extrapolation to zero defines 
the nonthermal critical point on the $\Delta U$ axis. We find 
-- for both values of $U_0$ -- the critical amplitude $\Delta U
\simeq 0.133 U_0$. 
Testing different cutoff parameters confirmed that $t_0=15$ is enough to determine
$\tau^{-1}_\mathrm{AFM}$ within the numerical
accuracy of the nonequilibrium DMFT simulation, as
Fig.~\ref{fig:afm_nonthermal}(c)--(d) demonstrate. 
This illustrates the potential of the cutoff scheme in studies
of dynamical phase transitions and nonthermal dynamics.

\subsection{Holstein-Hubbard model}
 
\begin{figure}[t] 
\centering
\includegraphics[width=\columnwidth]{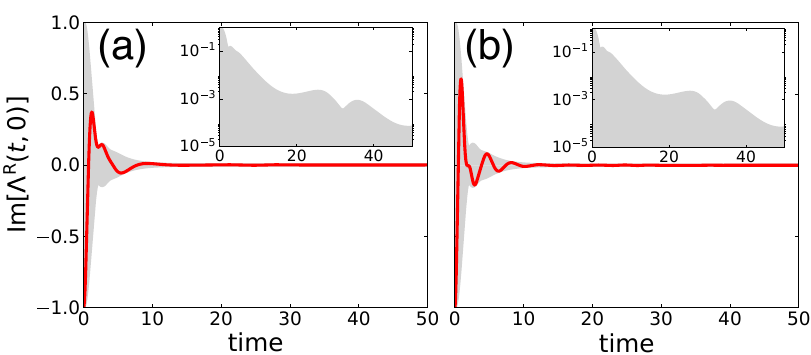}
\caption{Retarded component $\Lambda^\mathrm{R}_{\sigma}(t,0)$ of the
  hybridization function in equilibrium for (a) $U_0=4$, and (b)
  $U_0=6$. The gray shaded background represents the envelope function
  $|\Lambda^>_{\sigma}(t,0)|+|\Lambda^<_{\sigma}(t,0)|$ (on a logarithmic scale in the inset).
 \label{fig:langfirs_gret_decay} }
\end{figure}

\begin{figure*}[t] 
\centering
\includegraphics[width=0.9\textwidth]{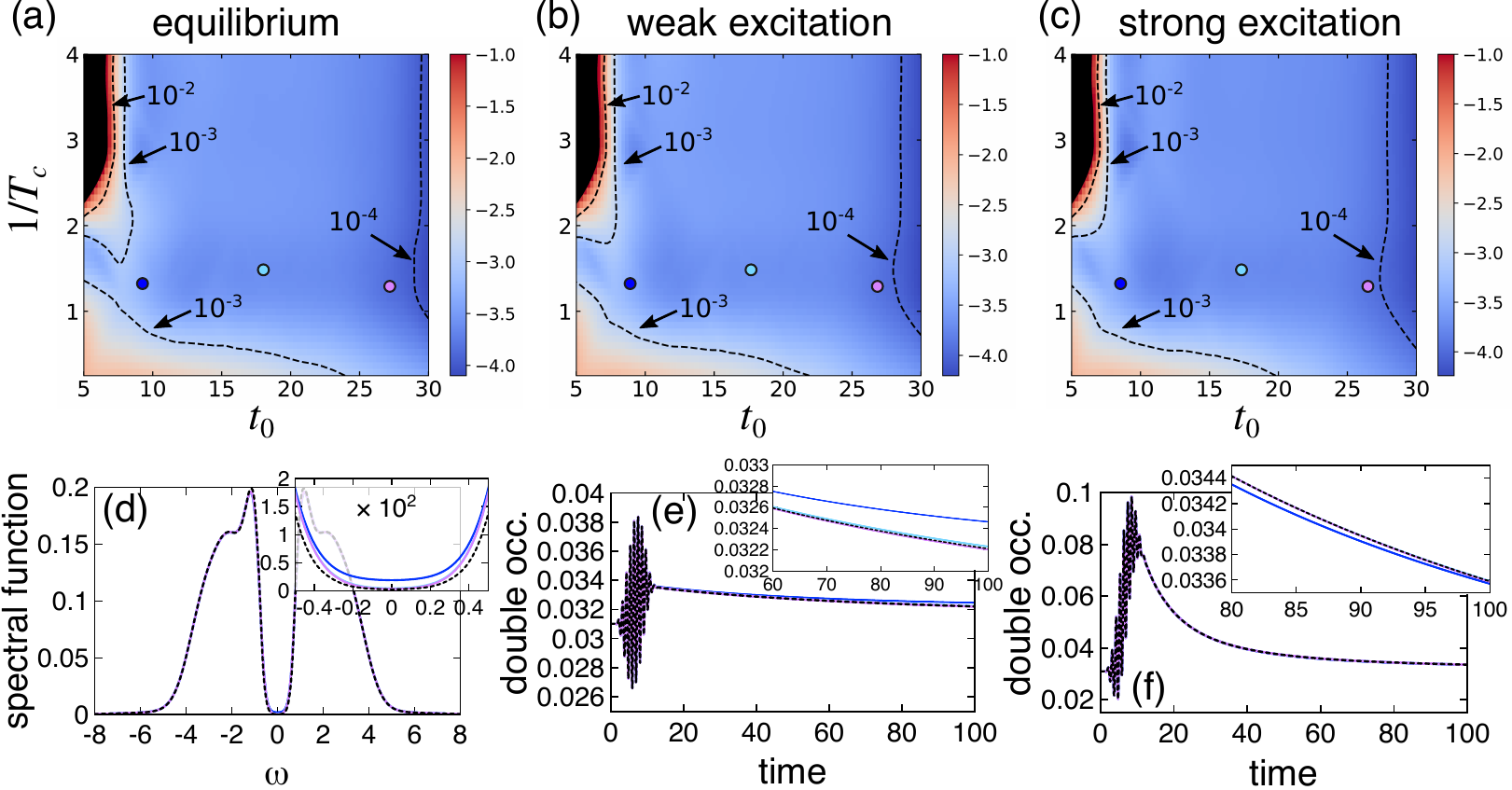}
\caption{(a)--(c): Norm distance (logarithm) between the two-time
  reference GF $G_{\mathrm{loc},\uparrow}(t,t^\prime)$ (no cutoff) and
  the GF obtained by the cutoff scheme for $U_0=4$ and $\Delta U=0$
  (a), $\Delta U =0.4$ (b) and $\Delta U=2.0$ (c). (d) Equilibrium
  spin-up spectral function from reference GF (black dashed) and
  cutoff scheme (colors as in (a)). The dynamics of the double
  occupation is shown in (e) for $\Delta U = 0.4$, and (f)
  $\Delta U=2.0$.  \label{fig:langfirs_u4}}

\end{figure*}

We now proceed to the case of the Hubbard-Holstein model,
which corresponds to the Hamiltonian~\eqref{eq:hubbhol} with $g > 0$. The
Hubbard-Holstein model can be treated within the framework of DMFT by
applying the Lang-Firsov
transformation~\cite{werner_phonon-enhanced_2013}, which (within NCA) maps the problem
to an effective impurity problem similar to that of the Hubbard model, up to an additional boson Green's function
which multiplies the hybridization function in the pseudo-particle Dyson equations. 
We have propagated the nonequilibrium DMFT
scheme up to $t_\mathrm{max}=100$, using a discretization of $\Delta t
=0.02$. As for the Hubbard model, we study a small-gap
insulator ($U_0=4.0$) and a large-gap system ($U_0=6.0$). The phonon
frequency is chosen as $\omega_\mathrm{ph}=0.2$, while the
e--ph coupling is fixed to $g=0.2$. Hence, the Lang-Firsov
parameter is $g/\omega_\mathrm{ph}=1$ while
the effective phonon coupling is 
$\lambda=g^2/\omega_\mathrm{ph} < 1$, justifying the
applicability of the NCA treatment using the Lang-Firsov
transformation.

Fig.~\ref{fig:langfirs_gret_decay} depicts the retarded hybridization function
along with its bounding function for the equilibrium case. One finds a
rapid decay which is even faster than for the paramagnetic Hubbard
model. This is understood by the additional broadening introduced by
the e--ph coupling, albeit phonon-induced oscillations lead
to a non-monotonic behavior. Truncating the memory is hence expected
to be a good approximation.

The results for $U_0=4$ are summarized in
Fig.~\ref{fig:langfirs_u4}. First one notices an unstable region for
$1/T_c \approx 2$ and $t_0<10$, in which the solution diverges. The
origin of this instability is similar to the AFM case: a sharp
cutoff ($T_c < 0.5$) of the oscillating hybridization function leads
to negative spectral weight, which indicates a violation of particle number
and spectral weight conservation. 
%conservation and the positivity of the spectral function.
%
Apart from this potentially unstable region of small cutoff times, the
error is of the order $10^{-3}$ for most values of $T_c$ and $t_0$,
but decreases very slowly with $t_0$. This behavior is very similar
for equilibrium (Fig.~\ref{fig:langfirs_u4}(a)), weak
(Fig.~\ref{fig:langfirs_u4}(b)) and strong excitation
(Fig.~\ref{fig:langfirs_u4}(c)). In contrast to the paramagnetic
Hubbard model, the e--ph coupling leads to a rapid dissipation of
kinetic energy, i.e., an efficiently cooling the photoexcited
doublons. For this reason, the filling of the gap is far less
pronounced than for the pure Hubbard model, and the spectral functions
of the excited system are very similar to the equilibrium case.

Nevertheless, local observables like the double occupation
(Fig.~\ref{fig:langfirs_u4}(e)--(f)) can be converged to match the
exact propagation up $\sim 10^{-5}$ by increasing the cutoff time to
$t_0\approx 17$. The cutoff temperature $1/T_c$ plays only a minor
role for large enough $t_0$, although there is an optimal region
around $T_c \approx 0.6$, which provides a good compromise between not
affecting the short-time dynamics and suppressing the hybridization
function for larger $|t-t^\prime|$ in a smooth way.

We have also analyzed the case $U_0=6$ (see Appendix), which is qualitatively similar
to $U_0=4$. The error of the spectral functions and especially of the
double occupation is, however, reduced by one order of magnitude. This
is analogous to the observations for the Hubbard model
(Fig.~\ref{fig:sbhb_u6}). Moreover, there is no unstable region in
this case.

\begin{figure}[t] 
\centering
\includegraphics[width=\columnwidth]{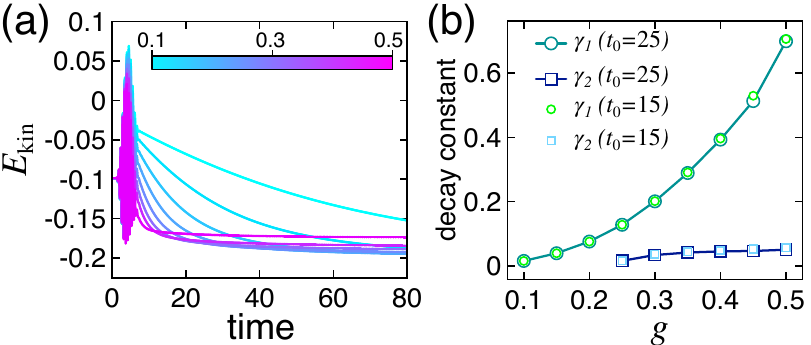}
\caption{(a) Dynamics of the kinetic energy for $U_0=10$, $\Delta
  U=5$, $\omega_\mathrm{ph} = 1$ and $g$ ranging from $0.1$ to $0.5$
  (color scale). (b) The two expontial decay constants of the kinetic
  energy observed in (a), for cutoff time $t_0=25$ (large symbols and
  lines) and $t_0=15$ (smaller symbols).
\label{fig:langfirs_ekin_decay}}
\end{figure}

We have further analyzed the cooling dynamics induced by e--ph coupling 
and the ability of the cutoff scheme to reproduce the corresponding timescales. 
In these calculations we use a larger Hubbard repulsion $U_0=10$ to exclude impact
ionization processes and heating by doublon-holon recombination. 
Without e--ph coupling, the kinetic energy and
the double occupation stay effectively constant. In contrast,
including e--ph interactions, the photoexcited doublons can
efficiently dissipate their kinetic energy $E_\mathrm{kin}$ as long as it is larger
than the phonon energy $\omega_\mathrm{ph}$. A further decrease of
$E_\mathrm{kin}$ becomes inefficient since no full quanta of phonon
excitations can be emitted (so-called phonon bottleneck~\cite{sentef_examining_2013,kemper_effect_2014}). Therefore,
two time scales for the relaxation of $E_\mathrm{kin}$ are
expected. Figure~\ref{fig:langfirs_ekin_decay}(a) shows the dynamics of the
kinetic energy after a strong pulse excitation ($\Delta U = 0.5 U_0$)
for $t_0=25$ and $T_c=0.75$. Comparing with the reference solution,
one finds a difference of at most $10^{-4}$. Analyzing the relaxation
dynamics, a single-exponential decay is found for a small magnitude of
the e--ph coupling $g$, while two different decay regimes become
apparent for larger $g$. A double-exponential fit yields the
decay constants $\gamma_1$ (fast relaxation) and $\gamma_2$ (slow
relaxation), presented in
Fig.~\ref{fig:langfirs_ekin_decay}(b). Consistent with
Ref.~\onlinecite{werner_field-induced_2015}, one finds a quadratic
scaling of $\gamma_1$ with $g$, while the slow relaxation described by
$\gamma_2$ does not increase significantly with $g$ for
$g>0.3$. Comparing the decay constants for the smaller value of the
memory time $t_0=15$ and the larger value $t_0=25$, small deviations
are visible for larger $g$. Besides these small differences,
$\gamma_1$ and $\gamma_2$ are well reproduced even by
$t_0=15$. Comparing with the reference solution we found the error of
$\gamma_{1,2}$ to be less than $10^{-4}$ for $t_0=25$.

\section{Conclusions\label{sec:conclusions}}

We studied the effect of truncations of the memory kernel in
Kadanoff-Baym equations, with a focus on nonequilibrium DMFT
simulations. These calculations are based on the solution of lattice
and impurity Dyson equations in which a lattice, impurity or
pseudo-particle self-energy plays the role of a memory kernel. The
standard techniques for the numerical solution of these equations
involve a discretization of the time interval of length $t_\text{max}$
into $N_t$ time slices of length $\Delta t$ and their computational
cost (for given self-energy) scales as $N_t^3$. If the memory time of
the self-energy is truncated at some $t_\text{cut}$ (corresponding to
$N_c$ time-slices), the computational effort can be reduced to
$\mathcal{O}(N_c N_t^2)$. 
A further reduction to $\mathcal{O}(N_c^2 N_t)$ is possible if the two-time GF is only calculated
on $N_c$ time slices, e. g. because one is only interested in time-local observables. 
The appealing feature of the self-energy truncation approach is its simplicity,
and that in addition to the speed-up in the calculations, it also enables
a potentially significant reduction in the memory requirements.    
An alternative route to overcome the
memory bottleneck could be the use of a suitable compressed representation of the
two-time self-energy~\cite{balzer_auxiliary_2014}, but such schemes
are difficult to implement in a controlled way.

Considering typical parameter choices for nonequilibrium DMFT
simulations of the Hubbard model in the strongly interacting regime,
we have found that a memory time of less than 10 inverse hoppings is
fully adequate to describe the time evolution of the system after a
strong perturbation. In the case of weak perturbations,
e. g. photo-doping concentrations of less than one percent, memory
times of 20-30 inverse hoppings are needed to reduce the truncation
error to a negligible level. (In practice we found that a norm error
of the Green's function $\lesssim 10^{-4}$ results in negligible
effects on relevant observables). As expected, the longest memory
times are found in situations where the (nonequilibrium) spectral
function exhibits sharp features, as is the case in the
antiferromagnetic Mott state, or in equilibrium Mott insulators at low
temperature, which feature sharp band edges. The coupling to phonons
%can, at least in the Lang-Firsov approach considered here, also lead to slowly decaying
%oscillations in the memory kernel, which means that relatively long memory time
%of the order of 25 inverse hoppings are needed for accurate results. 
results %, at least in the Lang-Firsov approach considered here, 
in a slower convergence 
to the exact result with $t_\text{cut}$, especially in the strongly excited regime, 
because the system is cooled down by the phonons and the original gapped spectrum 
is recovered in the long-time limit. 
While a memory time of 10 yields a  
norm error in the Green's function of $10^{-3}$, $t_\text{cut}\approx 30$ is needed to reduce 
this error to $10^{-4}$. 

While one might naively expect that smooth cutoffs of the memory time
reduce artifacts, we found that the sharpness of the cutoff has little
effect on the quality of the approximation in the paramagnetic case, even though in the
equilibrium and weakly perturbed systems a nontrivial optimal ``cutoff
temperature" of the order of $T_c\approx 0.5-0.75$ could be
identified. In the antiferromagnetic %and electron-phonon coupled systems, 
system, 
where the 
hybridization function decays more slowly, the sharpness of the cutoff matters, but for
a large enough cutoff time, the same cutoff temperatures as in the paramagnetic 
Hubbard model are adequate. 

We have demonstrated that simulations with memory cutoff correctly
reproduce the different characteristic timescales appearing in the
nonequilibrium evolution of photoexcited strongly-correlated lattice
systems. Not surprisingly, this is true for fast processes, such as
the ``photo-doping" by the $U$-modulation, the generation of
additional doublon-holon pairs by impact ionization, or the rapid
cooling of the photo-carriers in the presence of strong
electron-phonon coupling. These occur on the timescale of a few
inverse hoppings, which is comparable to the memory cutoff time. More
remarkable is the fact that also the slower processes, in particular
the thermalization timescale associated with doublon-holon
production or recombination or the slow melting of AFM order near the nonthermal
critical point, are accurately captured by the cutoff scheme. This
suggests that under generic conditions, we can correctly reproduce the
full nonequilibrium DMFT dynamics, including all the relevant fast and
slow processes, as well as transient trapping phenomena, using a
numerically efficient memory truncation scheme with
$t_\text{cut}\ll t_\text{max}$.

The short memory times $t_\mathrm{cut} \lesssim 10$ in strongly
excited systems furthermore imply huge potential efficiency gains in
simulations based on higher-order perturbative impurity solvers. For
instance, the fourth order weak-coupling expansion or the OCA-level
strong coupling expansion scales as $N_t^4$. A truncation of the
convolution integrals in these calculations reduces the computational
effort to $\mathcal{O}(N_c^2 N_t^2)$. For $t_\text{cut}\ll t_\text{max}$ the
truncation results in a significant speed-up of the simulations and it
is thus useful to explore under which conditions the errors incurred
by the truncation remain negligible. For higher-order treatments, the
cost of the evaluation of the self-energy can be reduced from
$\mathcal{O}(N_t^n)$ (with relatively large $n$) to $\mathcal{O}(N_t^2 N_c^{(n-2)})$. In
this context, it is interesting to note that a Monte Carlo based
scheme such as the inchworm algorithm~\cite{cohen_greens_2014}
automatically exploits these short memory times, which enables the
sampling up to fairly high orders. In fact, the inchworm approach is
based on the same step-wise time propagation and the same renormalized
strong-coupling pseudo-particle propagators as the perturbative
strong-coupling methods \cite{eckstein_nonequilibrium_2010}, and it
would be interesting to compare the computational efficiency of the
Monte Carlo sampling approach with that of a self-energy evaluation
based on the truncation scheme introduced here. However, since the
implementation of the perturbative strong-coupling diagrams beyond the
third order requires a substantial coding effort, we leave this as an
interesting topic for future investigations. 
%Furthermore, we note that
%the $\mathcal{O}(N^3_t)$ scaling (or higher) could be further reduced to
%$\mathcal{O}(N^2_c N_t)$ in case the two-time GFs are only computed
%for $N_c$ time slices. Such an approach would be particularly useful if
%one is primarily interested in single-particle observables, which are
%obtained from the time-diagonal GF.

Another promising application of our scheme is its use within the
Floquet-implementation of
DMFT~\cite{tsuji_correlated_2008,mikami_brillouin-wigner_2016,
murakami_nonequilibrium_2017,qin_spectral_2017,sorantin_impact_2018},
which directly treats the nonequilibrium steady state of periodically
driven systems. These calculations include a coupling to a heat bath
and interesting applications such as the high-harmonic generation in
solids~\cite{murakami_high-harmonic_2017} involve the simulation of
highly excited nonequilibrium states. The memory times in these
simulations can be expected to be short, and the cutoff scheme
introduced here will enable more efficient simulations at low driving
frequency, or the use of more accurate higher-order impurity solvers.

\acknowledgements

This work was supported by the Swiss National Science Foundation
through NCCR MARVEL and the European Research Council through ERC
Consolidator Grant 724103. The calculations have been performed on the
Beo04 cluster at the University of Fribourg, and the Piz Daint cluster
at the Swiss National Supercomputing Centre (CSCS).

\appendix

\section{Results for larger gap size}

In this appendix, we present the norm error, spectral function and
time evolution of the double occupation or magnetization for the
Hubbard and Holstein-Hubbard model with larger gap ($U_0=6$).

\begin{figure*}[t] 
\centering
\includegraphics[width=1\textwidth]{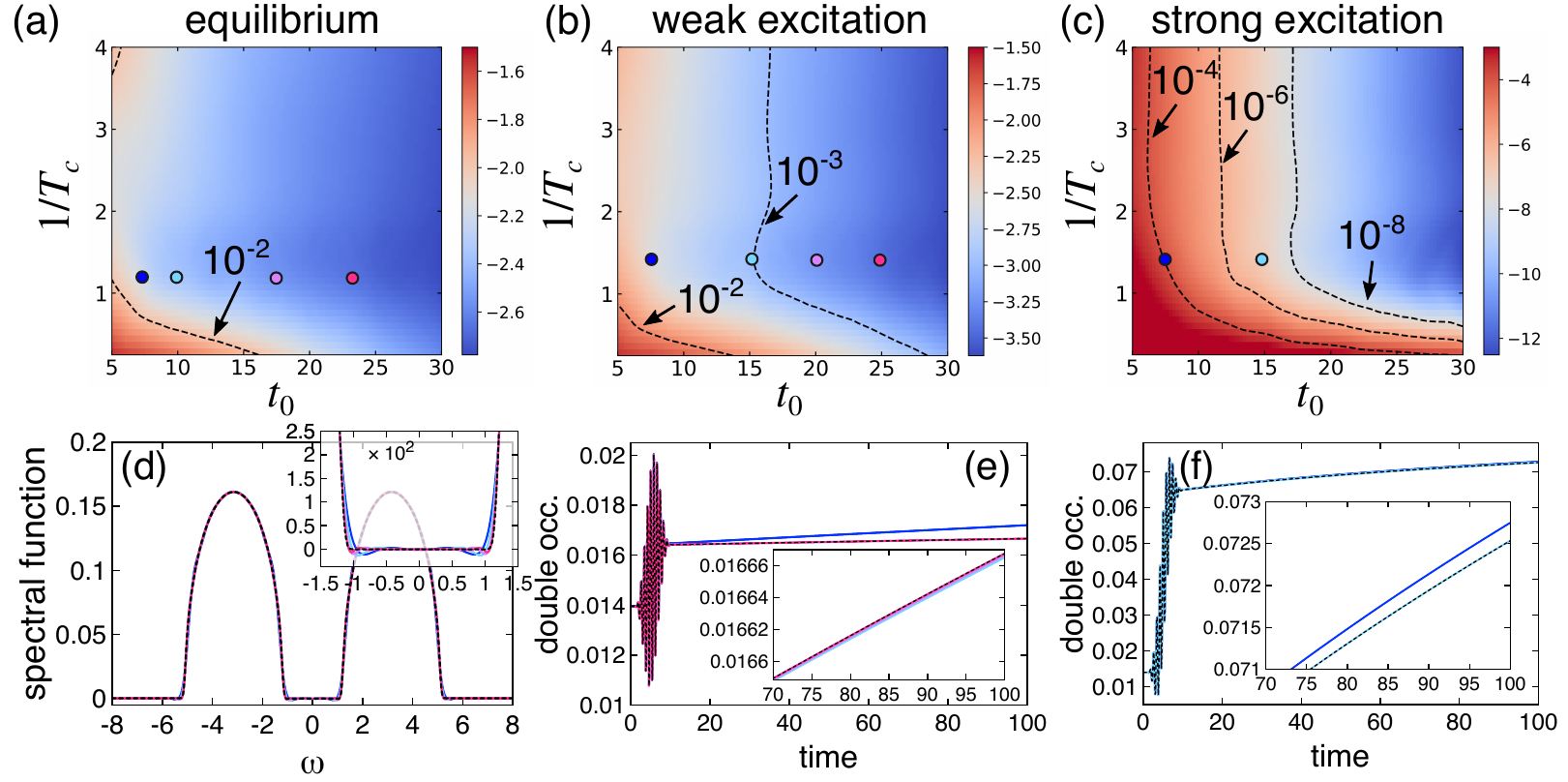}
\caption{
  Hubbard model in the paramagnetic phase. 
  (a)--(c): Norm distance (logarithm) between the two-time reference GF (no cutoff)
  and $G_\mathrm{cut}$ obtained by the cutoff scheme as a function of the memory
  time $t_0$ and sharpness of the cutoff $1/T_c$ (cf.~\eqref{eq:fermicut})
  for $U_0=6$ and $\Delta U=0$ (a), $\Delta U =0.6$ (b) and $\Delta
  U=3.0$ (c). The contour lines delimit the regions where the error
  is smaller than the given values. The colored dots indicate 
  representative values of $t_0$ and $T_c$, for which the
  equilibrium spectral function (d) and the double
  occupancy for weak (e) and strong excitation (d) are
  shown (consistent color coding). The black dashed lines represent
  the reference results.
  }
\label{fig:sbhb_u6}
\end{figure*}

\begin{figure*}[t] 
\centering
\includegraphics[width=1\textwidth]{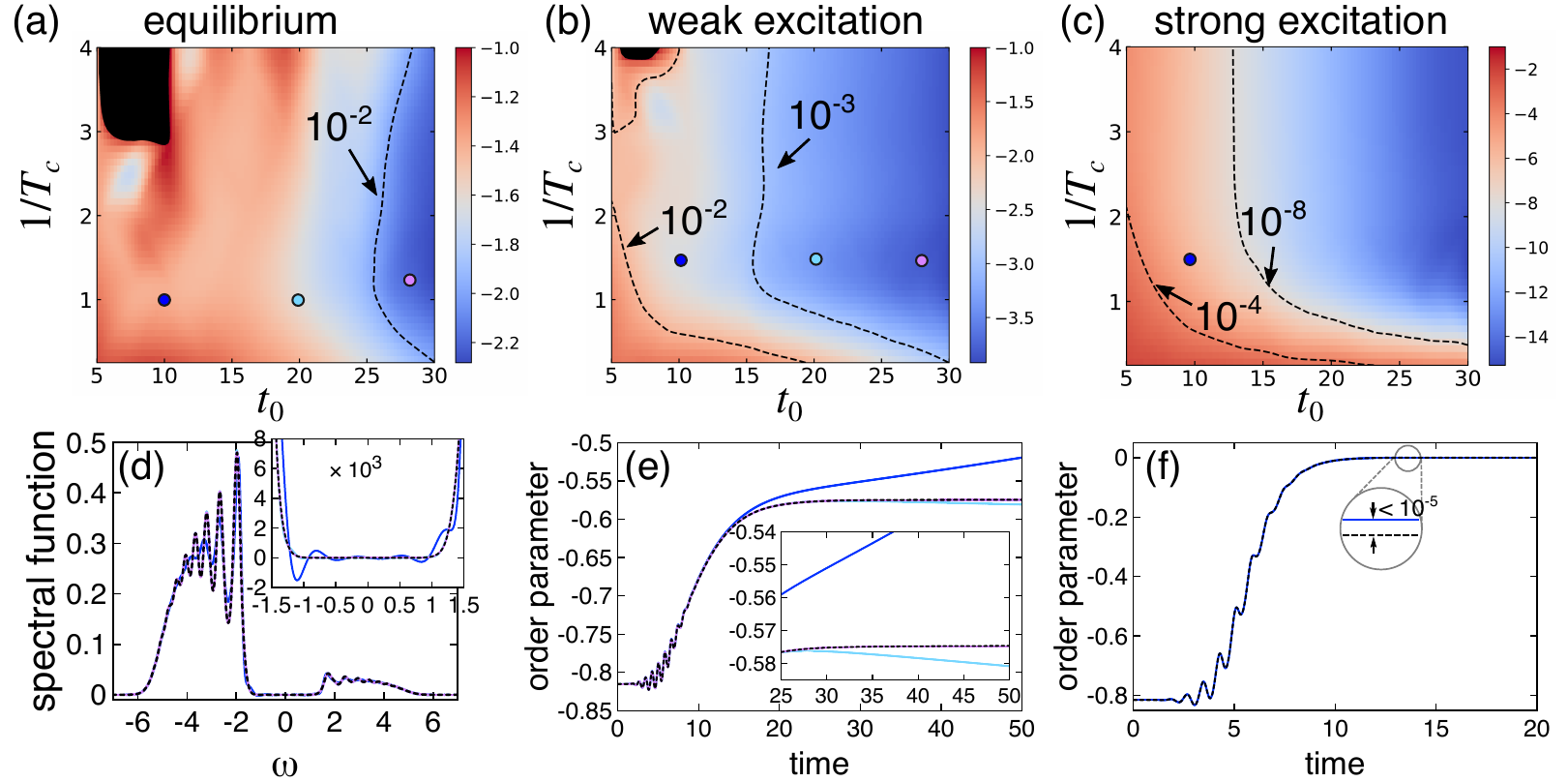}
\caption{
  Hubbard model in the antiferromagnetic phase. 
  (a)--(c): Norm distance (logarithm) between the two-time
  reference GF $G_{\mathrm{loc},\uparrow}(t,t^\prime)$ (no cutoff) and
  the GF obtained by the cutoff scheme, analogous to
  Fig.~\ref{fig:afm_u4}, for $U_0=6$ and $\Delta U=0$ (a),
  $\Delta U =0.6$ (b) and $\Delta U=3.0$ (c). (d) Equilibrium spin-up spectral function from reference
  GF (black dashed) and cutoff scheme (colors as in (a)). The dynamics
  of the AFM order parameter
 is shown in (e) for $\Delta U = 0.6$, and (f)
  $\Delta U=3.0$.  }
\label{fig:afm_u6}
\end{figure*}

\begin{figure*}[t] 
\centering
\includegraphics[width=1\textwidth]{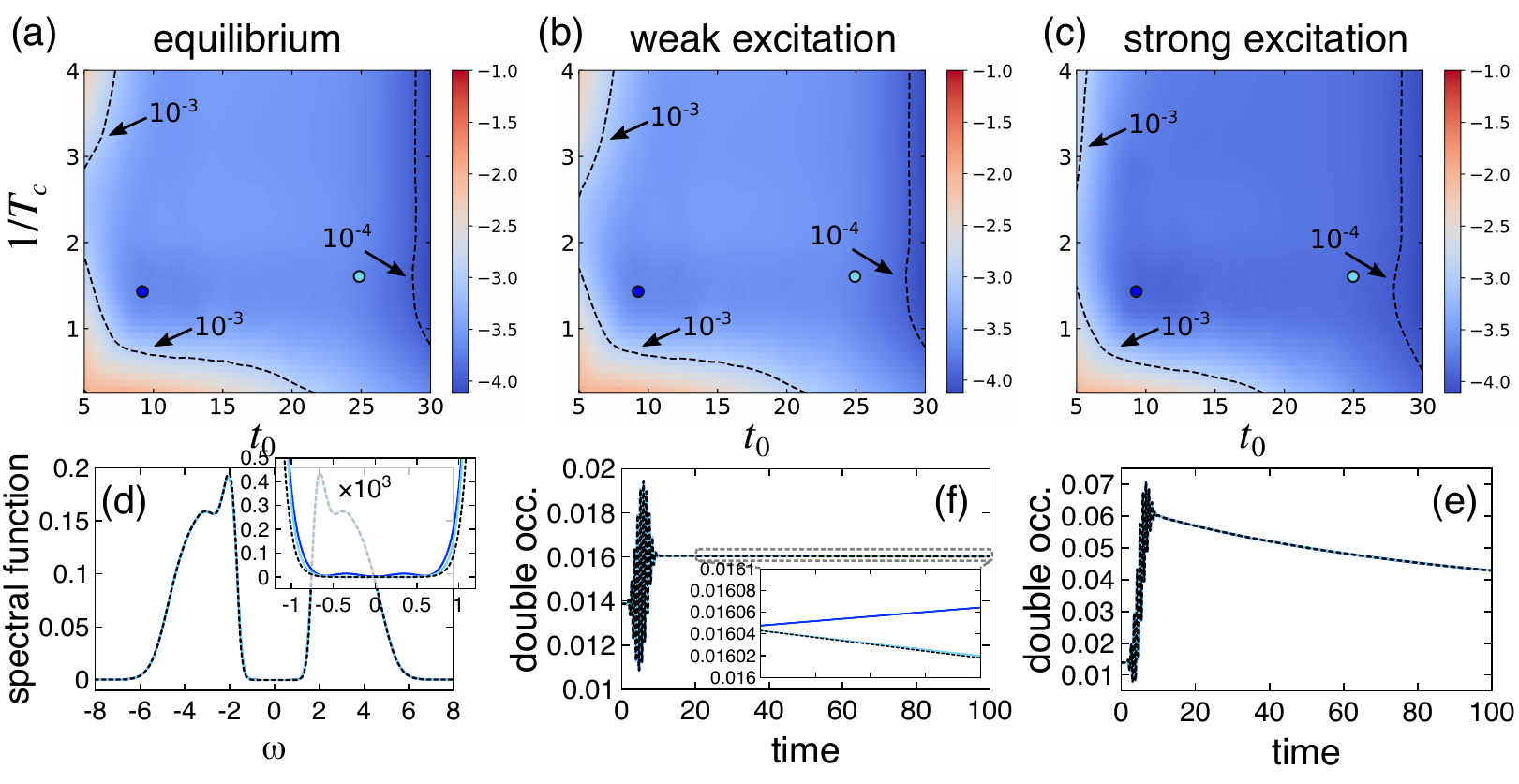}
\caption{Holstein-Hubbard model in the paramagnetic phase ($g=0.2$, $\omega_0=0.2$). (a)--(c): Norm distance (logarithm) between the two-time
  reference GF $G_{\mathrm{loc},\uparrow}(t,t^\prime)$ (no cutoff) and
  the GF obtained by the cutoff scheme for $U_0=6$ and $\Delta U=0$
  (a), $\Delta U =0.6$ (b) and $\Delta U=3.0$ (c). (d) Equilibrium
  spin-up spectral function from reference GF (black dashed) and
  cutoff scheme (colors as in (a)). The dynamics of the double
  occupation is shown in (e) for $\Delta U = 0.6$, and (f)
  $\Delta U=3.0$.  }
\label{fig:langfirs_u6}
\end{figure*}

\end{document}